\documentclass[aps,prl,twocolumn,amsmath,amssymb,floatfix,superscriptaddress,10pt]{revtex4-2}
\usepackage{blindtext}
\usepackage{epsfig}
\usepackage{color}
\usepackage{amsmath}
\usepackage{amsfonts}
\usepackage{amssymb}
\usepackage{graphicx}%
\usepackage{url}
\usepackage[breaklinks]{hyperref}
\usepackage[english]{babel}
\usepackage{esvect}
\usepackage[normalem]{ulem}
\usepackage[inline]{enumitem}
\usepackage{xcolor}
\usepackage{xfrac}
\usepackage{dsfont}
\usepackage{bm}
\usepackage{placeins}
\usepackage{array}
\newcolumntype{C}[1]{>{\centering\let\newline\\\arraybackslash\hspace{0pt}}m{#1}}
\usepackage{footnote}
\usepackage{isomath}
\DeclareMathAlphabet\mathbfcal{OMS}{cmsy}{b}{n}
\usepackage{esint} 

\hypersetup{colorlinks,citecolor=blue,filecolor=blue,linkcolor=blue,urlcolor=blue}

{
 \definecolor{BLACK}{gray}{0}
 \definecolor{WHITE}{gray}{1}
 \definecolor{RED}{rgb}{1,0,0}
 \definecolor{GREEN}{rgb}{0,1,0}
 \definecolor{BLUE}{rgb}{0,0,1}
 \definecolor{CYAN}{cmyk}{1,0,0,0}
 \definecolor{MAGENTA}{cmyk}{0,1,0,0}
 \definecolor{YELLOW}{cmyk}{0,0,1,0}
}

\begin{document} 

\title{Theory of local orbital magnetization: local Berry curvature}
\author{Sariah Al Saati}
\email{sariah.alsaati@gmail.com}
\affiliation{CPHT, CNRS, École polytechnique, Institut Polytechnique de Paris, 91120 Palaiseau, France}

\author{Karyn Le Hur}
\email{karyn.le-hur@polytechnique.edu}
\affiliation{CPHT, CNRS, École polytechnique, Institut Polytechnique de Paris, 91120 Palaiseau, France}

\author{Frédéric Piéchon}
\email{frederic.piechon@universite-paris-saclay.fr}
\affiliation{Laboratoire de Physique des Solides, Université Paris-Saclay, CNRS, 91405 Orsay, France}

\begin{abstract}
We develop a thermodynamic theory of local orbital magnetization based on a perturbative expansion of the magnetic-field-dependent local density of states. Applicable to periodic crystals, finite systems with open boundaries, and ribbons alike, the formalism resolves orbital magnetic textures at the sublattice scale. It further reveals a previously unidentified local Berry curvature that captures the magnetic-field-induced redistribution of electronic weight in real space. We establish the consistency of the theory across geometries, identify orbital ferro-, antiferro-, and ferrimagnetic phases in both topological and trivial insulators, and demonstrate that the local Berry curvature provides a bulk description of topology in finite systems.
\end{abstract}
\maketitle

Magnetism is one of the most familiar  properties of materials; easily observed, known since antiquity, and central to modern technologies. Magnetization in a material arises from two sources: the spin of the electron and its orbital motion.  Spin magnetization is tied to collective behavior, giving rise to various kinds of magnetic order (ferromagnetic, antiferromagnetic, etc.).  By contrast, orbital magnetization is usually understood at the single-particle level as the average orbital momentum of electrons at thermodynamic equilibrium. Despite this apparent simplicity, it is only since 2005 that a consistent quantum theory of orbital magnetization for crystalline solids has been established \cite{Thonhauser2005,Xiao2005,Ceresoli2006,Shi2007,Xiao2010,Thonhauser2011}.

In this modern theory, orbital magnetization in crystals is defined as
\begin{align}
{\bm M}_{\textrm{orb}} (\mu, T)&=  \dfrac{e}{\hbar}   \sum_{n} \left\langle  f(\varepsilon_{n\bm k}){\bm m}_{n \bm k} + F(\varepsilon_{n \bm k}) {\bm \Omega}_{n \bm k} \right\rangle_{\bm k}
    \label{eq:moderntheory}
\end{align}
where ${\bm m}_{n \bm k}$ and ${\bm \Omega}_{n \bm k}$ are, respectively, the orbital momentum and the Berry curvature of the Bloch state of energy $\varepsilon_{n \bm k}$;  $f(\varepsilon) = 1/(1+e^{\beta(\varepsilon-\mu)})$ and $F(\varepsilon) = k_B T \ln\left(1 + e^{-\beta(\varepsilon - \mu)}\right)$ reflect Fermi statistics,  $\mu$ is the chemical potential, and $\beta = 1/(k_BT)$ is the inverse temperature. We use the notation $\langle \bullet \rangle_{\bm k} =\int \frac{d^dk}{(2\pi)^d} \bullet$.
A major shift brought about by this expression was the realization that Berry curvature constitutes a significant source of orbital magnetization alongside the conventional orbital magnetic moment; where both quantities encode virtual interband effects.

From a thermodynamic point of view, the orbital magnetization is obtained as the zero field limit of the first derivative ${\bm M}(\mu)=-\partial_{\bm B} \Xi|_{\bm B=0}$ of the grand potential  $\Xi(\mu, {\bm B})= -\int\mathrm{d}\varepsilon F(\varepsilon) \rho(\varepsilon,{\bm B})$. In this framework, the key quantity is the field dependent density of states $\rho(\varepsilon,{\bm B})$. To retrieve the modern theory (\ref{eq:moderntheory}), it takes the effective form
\begin{align}
    \rho(\varepsilon,\bm B) =\sum_{n} 
    & \left\langle \left(1+\frac{e}{\hbar}{\bm \Omega}_{n \bm k}\cdot \bm B \right) \right. \nonumber \times \\
    & \left. \delta\left(\varepsilon-\varepsilon_{n \bm k} -\frac{e}{\hbar}{\bm m}_{n \bm k}\cdot\bm B \right) \right\rangle _{\bm k} .
    \label{eq:dosk}
\end{align} 
In this picture, as expected, the orbital moment ${\bm m}_{n\bm k}$ quantifies the field induced correction of the energy band $\varepsilon_{n\bm k}$. Instead, the Berry curvature ${\bm \Omega}_{n\bm k}$  quantifies an effective field induced spectral weight transfer between bands \cite{Xiao2005,Xiao2010}.

A key property of orbital magnetization, made explicit by modern theory, is its close connection to Hall insulators through the St\v{r}eda formula $\partial_\mu M_{\text{orb}} = \frac{1}{e}\sigma^{xy}$ \cite{Streda1982}, which relates the variation of orbital magnetization with chemical potential to Hall conductivity. In an insulating gap, the quantized Hall conductivity plateau $\sigma^{xy} = \frac{e^2}{h}C$, is determined by the Chern number $C = 2\pi \sum_n\left\langle f(\varepsilon_{n\bm k}){ \Omega}_{n \bm k} \right\rangle _{\bm k}$ of the occupied bands, implying that the slope of the orbital magnetization is itself a topological quantity. Since the Středa formula remains valid beyond periodic crystals, it naturally raises the question of how topology is encoded locally in orbital magnetization in finite and noncrystalline systems. Although the local Chern marker introduced by Bianco and Resta provides a real-space characterization of topology \cite{Bianco2011,Bianco2013}, a microscopic theory of local orbital magnetization valid across periodic crystals, finite systems, and ribbon geometries has remained absent.

In this Letter, we develop such a theory from the magnetic-field dependence of the local density of states. The resulting thermodynamic formalism provides a unified description of local orbital magnetization across all geometries while resolving orbital magnetic textures at the single-site and sublattice scales. It further uncovers a previously unidentified local Berry curvature, which describes the magnetic field-induced redistribution of electronic weight in real space. We demonstrate that this quantity provides a local bulk marker of topology in finite systems, thereby extending the modern theory of orbital magnetization into real space.

{\color{blue}{
\textit{Thermodynamic theory of local orbital magnetization-}}}
We formulate a local thermodynamic theory by extending the modern theory of orbital magnetization to the local grand potential. Specifically, the local orbital magnetization is defined as the zero-field limit
${\bm M}({\bm r},\mu)=-\partial_{\bm B} \Xi({\bm r})|_{\bm B=0}$
where  $\Xi(\bm r,\mu, {\bm B})= -\int\mathrm{d}\varepsilon F(\varepsilon) \rho({\bm r},\varepsilon,{\bm B})$, is the local grand potential, in which the central quantity is the magnetic field dependent local density of states (LDOS) $\rho(\bm r, \varepsilon,{\bm B})$, more precisely its first order field induced correction.

We consider noninteracting spinless electrons in a tight-binding model, where the magnetic field is incorporated through the Peierls substitution \cite{Peierls1933}, and evaluate the LDOS within the gauge-invariant perturbation theory of Ref. \cite{Raoux2015}. 
The resulting local {\it onsite} orbital magnetization reads:
\begin{align}
\hspace{-.3cm}
{\bm M}_{\textrm{orb}}({\bm r},\mu) &=\dfrac{e}{2\pi \hbar} \int\mathrm{d}\varepsilon F(\varepsilon) \mathrm{Re} \langle \bm r|  {\cal G}[(\hat {\bm v} {\cal G})\times (\hat {\bm v} {\cal G})] | \bm r \rangle 
\label{eq:magldos}
\end{align}
where ${\cal G}(\varepsilon)=(\varepsilon-{\cal H})^{-1}$ is the retarded single particle Green's function associated to the zero (magnetic) field Hamiltonian ${\cal H}$, and $\hat {\bm v}=-i [\hat {\bm r},{\cal H}]$ is the zero field velocity operator (in unit of $\hbar$). 
In what follows, we show how \eqref{eq:magldos} naturally gives rise to distinct yet fully consistent descriptions of local orbital magnetization in finite systems, periodic crystals, and ribbons like geometries.

{\color{blue} \textit{Onsite magnetization in open-boundary finite systems-}}

We first consider the site-resolved local orbital magnetization of a finite-size system with open boundaries, the most general setting encompassing molecules, crystalline flakes,  disordered samples, and amorphous structures.
Evaluating Eq. \eqref{eq:magldos} yields (see supplementary material \cite{supplementary})
\begin{align}
    {\bm M}_{\textrm{orb}}({\bm r}) = \dfrac{e}{\hbar} \sum_n \left[f(\varepsilon_n)  {\bm m}_{n}(\bm r)  + F(\varepsilon_n) {\bm \Omega}_n(\bm r) \right]
    \label{eq:localMorbFin}
\end{align} with 
\begin{align}
    {\bm m}_{n}(\bm r)&=  |\langle \bm r|n\rangle|^2 {\bm m}_{n},\label{eq:LocalBerry}\\ 
\bm \Omega_n(\bm r) &= \mathrm{Re}  \sum_{m \ne n} \langle \bm r|n\rangle \langle m|\bm r \rangle\dfrac{   \langle n| {\hat {\bm r}}\times  {\hat {\bm v}}|m\rangle}{\varepsilon_n - \varepsilon_m}   \label{eq:LocalBerry1}
\end{align} 
where $|n\rangle$ represents the finite-size eigenstate of energy $\varepsilon_n$ (at zero magnetic field), ${\bm m}_n= -\frac{1}{2} \langle n| {\hat {\bm r}}\times  {\hat {\bm v}}|n\rangle$ is the orbital momentum of state $|n\rangle$ (in the open-boundary finite geometry), and with ${\hat {\bm v}}=-i[\hat {\bm r},{\cal H}_{r}]$ the velocity and $\hat {\bm r}$ the position operators in $\bm r$-space representation (in practice ${\cal H}_{r}$ is a matrix $N\times N$  with $N$ the total number of sites, assuming a single orbital per site). 

Eq. \eqref{eq:localMorbFin} constitutes the real-space counterpart of the modern theory Eq. \eqref{eq:moderntheory}. The first term ${\bm m}_{n}(\bm r)= |\langle \bm r|n\rangle|^2 {\bm m}_{n}$ is simply the projection of orbital momentum ${\bm m}_n$ in real-space (onsite). The second term $\bm \Omega_n(\bm r)$ defines a previously unidentified local Berry curvature. Its emergence constitutes the central conceptual result of this work.

The physical meaning of these two contributions becomes transparent from the effective low-field LDOS,
\begin{align}
    \rho(\varepsilon, \bm r, \bm B) = \sum_{n} &\left(|\langle \bm r|n\rangle|^2 +\dfrac{e}{\hbar}\bm \Omega_{n } (\bm r) \cdot \bm B\right) \times \nonumber \\ &\delta\left(\varepsilon - \varepsilon_{n} - \dfrac{e}{\hbar} \bm m_{n}\cdot \bm B\right),
\end{align}
that naturally recovers  Eq. \eqref{eq:localMorbFin}.
As in the modern theory, the orbital moment $\bm m_{n}$ 
 describes the magnetic-field-induced shift of the energy level $\varepsilon_{n}$ ; importantly, this contribution is independent of position. By contrast, the local Berry curvature governs the field-induced correction to the local spectral weight $|\langle \bm r | n \rangle|^2$. Because it satisfies the sum rule $\sum_{\bm r} \Omega_{n } (\bm r) = 0$, it describes a redistribution of the electronic weight of eigenstate $|n\rangle$ in real space, without changing its normalization. Summing Eq.  \eqref{eq:localMorbFin} over all sites immediately eliminates the local Berry-curvature contribution through the above sum rule, yielding
\begin{align}
    {\bm M}_{\textrm{orb}} = \dfrac{e}{\hbar} \sum_n f(\varepsilon_n)  {\bm m}_{n}
    \label{eq:MorbTotrspace}
\end{align} which coincides with the conventional expression of total orbital magnetization in finite systems \cite{Thonhauser2005,Bianco2013}. Thus, the local Berry curvature leaves the total orbital magnetization unchanged. This implies that for the total finite system, the topological information is retrieved from the orbital momentum $\bm m_n$ of the occupied states.

{\color{blue}{\textit{Sublattice magnetization in periodic crystals-}}} 
We now consider the sublattice resolved local orbital magnetization of an infinite periodic system. The Bloch Hamiltonian ${\cal H}_{\bm k}$ acts in the $N_s$ dimensional sublattice space of the unit cell, giving rise to $N_s$ energy bands $\varepsilon_{n\bm k}$ with corresponding Bloch states $|n\bm k\rangle$. Denoting the sublattice positions ${\bm r}_{\alpha}$ with $\alpha=1,...,N_s$, \eqref{eq:magldos} yields the sublattice orbital magnetization (see supplementary material \cite{supplementary}) 
\begin{align}
    \hspace{-.2cm}
        {\bm M}_{\textrm{orb}}(\bm r_\alpha) &=  \dfrac{e}{\hbar}  \sum_{n}   \left\langle f(\varepsilon_{n\bm k}) {\bm m}_{n \bm k}(\bm r_\alpha) + F(\varepsilon_{n \bm k}) {\bm \Omega}_{n \bm k}(\bm r_\alpha) \right\rangle_{\bm k}
        \label{eq:MorbExtendedLocal}
    \end{align} 
    with
    \begin{align}
        \label{eq:MorbExtendedLocal2}
        {\bm m}_{n \bm k}(\bm r_\alpha) &= |\langle \bm r_\alpha| n \bm k \rangle|^2{\bm m}_{n \bm k} \\
        {\bm \Omega}_{n \bm k}(\bm r_\alpha) &= |\langle \bm r_\alpha| n \bm k \rangle|^2{\bm \Omega}_{n \bm k} + {\bm \Omega}^{\text{geom}}_{n \bm k}(\bm r_\alpha).
        \label{eq:MorbExtendedLocal3}
\end{align}
Equation \eqref{eq:MorbExtendedLocal} is the periodic-crystal counterpart of Eq. \eqref{eq:localMorbFin}. As in finite systems, the local orbital magnetization naturally separates into local orbital-moment and local Berry-curvature contributions. The local orbital moment is simply the projection of the Bloch orbital moment onto sublattice $\bm r_\alpha$, while the local Berry curvature contains two distinct terms. The first corresponds to the projection of the conventional Berry curvature onto the sublattice  $\bm r_\alpha$, whereas the second,$ {\bm \Omega}^{\text{geom}}_{n \bm k}(\bm r_\alpha)$
is a purely geometric contribution that has no analogue in the modern theory of the total orbital magnetization.
Here ${\bm m}_{n \bm k}=-\frac{1}{2}\sum_{m \ne n}\mathrm{Im} ({\bm r}_{nm \bm k} \times {\bm v}_{mn \bm k})$ and ${\bm \Omega}_{n\bm k}=\sum_{m \ne n}\mathrm{Im}({\bm r}_{nm \bm k} \times {\bm r}_{mn \bm k})$ denote the conventional orbital-momentum and Berry curvature of Bloch band $n$ respectively\cite{Berry}; where we have introduced ${\bm v}_{n m\bm k}=\langle n \bm k|\nabla_{\bm k} {\cal H}_{\bm k}|m \bm k \rangle$ the interband velocity and ${\bm r}_{n m\bm k}={\bm v}_{n m\bm k}/(\varepsilon_{n\bm k}-\varepsilon_{m\bm k})$  the interband position. 

For two band systems, the geometric contribution
${\bm \Omega}^{\text{geom}}_{n \bm k}(\bm r_\alpha)$ can be written explicitly as:
\begin{align}
    &{\bm \Omega}^{\text{geom}}_{n \bm k}(\bm r_\alpha)  = \dfrac{1}{2} |\langle \bm r_\alpha |n \bm k\rangle |^2  {\bm \Omega}_{n\bm k} - \dfrac{1}{2}  \sum_{m \ne n} |\langle \bm r_\alpha | m\bm k \rangle|^2 {\bm \Omega}_{n m\bm k} \nonumber \\
    &+\sum_{m \ne n} \mathrm{Im} \dfrac{\langle \bm r_\alpha| n \bm k \rangle \langle m \bm k| \bm r_\alpha\rangle}{\varepsilon_{n \bm k}-\varepsilon_{m \bm k}} {\bm r}_{nm \bm k}\times ({\bm v}_{n \bm k} + {\bm v}_{m \bm k}) 
\end{align} where ${\bm \Omega}_{n m\bm k}= \mathrm{Im}({\bm r}_{nm \bm k} \times {\bm r}_{mn \bm k})$ is the interband Berry curvature.  
The corresponding expression for multiband systems ($N_s \ge 3$) is given in the Supplemental Material \cite{supplementary}. Importantly, irrespective of the number of bands, the geometric contribution satisfies the sum rule
 $\sum_{\alpha} {\bm \Omega}^{\mathrm{geom}}_{n\bm k}(\bm r_\alpha) = 0$
which plays a central role in the following discussion.

The physical interpretation again follows from
the effective   LDOS,
\begin{align}
    \rho(\varepsilon, \bm r_\alpha, \bm B) = \sum_{n} &\left\langle \left(|\langle \bm r_\alpha|n\bm k\rangle|^2+\dfrac{e}{\hbar}\bm \Omega_{n \bm k} (\bm r_\alpha) \cdot \bm B\right) \right. \times \nonumber \\
    & \left. \delta\left(\varepsilon - \varepsilon_{n\bm k} - \dfrac{e}{\hbar}\bm m_{n\bm k}\cdot \bm B\right) \right\rangle _{\bm k}
    \label{eq:ldosK}
\end{align} 
which directly reproduces  Eq. \eqref{eq:MorbExtendedLocal}.
As in the modern theory, the orbital moment describes the magnetic-field-induced shift of the Bloch energy $\varepsilon_{n\bm k}$, while the local Berry curvature governs the magnetic-field-induced redistribution of the sublattice spectral weight $|\langle \bm r_\alpha|n\bm k\rangle|^2$.

Equation \eqref{eq:MorbExtendedLocal3} further reveals that this local Berry curvature naturally decomposes into two physically distinct contributions. The first term, $|\langle \bm r_\alpha|n\bm k\rangle|^2\,{\bm \Omega}_{n\bm k}$ is simply the projection of the usual Berry curvature ${\bm \Omega}_{n\bm k}$ onto sublattice $\bm r_\alpha$. Summing  over all sublattices, immediately recovers the conventional Berry curvature of Bloch band $n$, $\sum_\alpha |\langle \bm r_\alpha|n\bm k\rangle|^2 {\bm \Omega}_{n\bm k} = {\bm \Omega}_{n\bm k}$,
which itself satisfies the well-known sum rule $\sum_n {\bm \Omega}_{n\bm k}=0$.
 This contribution therefore describes the magnetic field induced transfer of the spectral weight 
 between different Bloch bands.

The second contribution,  ${\bm \Omega}^{\mathrm{geom}}_{n\bm k}(\bm r_\alpha)$ 
, instead redistributes the spectral weight of a given Bloch state among the different sublattices within the unit cell. Indeed, because
$\sum_{\alpha} {\bm \Omega}^{\mathrm{geom}}_{n\bm k}(\bm r_\alpha) = 0$, 
it preserves the total spectral weight of each Bloch state while modifying its internal spatial distribution. Consequently, summing Eq. \eqref{eq:MorbExtendedLocal} over all sublattices eliminates the geometric contribution and exactly recovers the modern theory [Eq. \eqref{eq:moderntheory}]. The latter therefore emerges as the unit-cell average of the present local formulation.

Finally, the same construction applies to ribbon geometries. Starting from Eq. \eqref{eq:magldos}, we obtain a local orbital magnetization that again separates into orbital-moment and local Berry-curvature contributions, thereby providing a unified description across finite systems,  periodic crystals and ribbon geometries. For brevity, the explicit expressions are given only in the Supplemental Material \cite{supplementary}. Ribbon geometries are nevertheless included throughout the numerical comparisons presented below.

\begin{figure}[ht!]
    \begin{center}
        \includegraphics[width=0.49\textwidth]{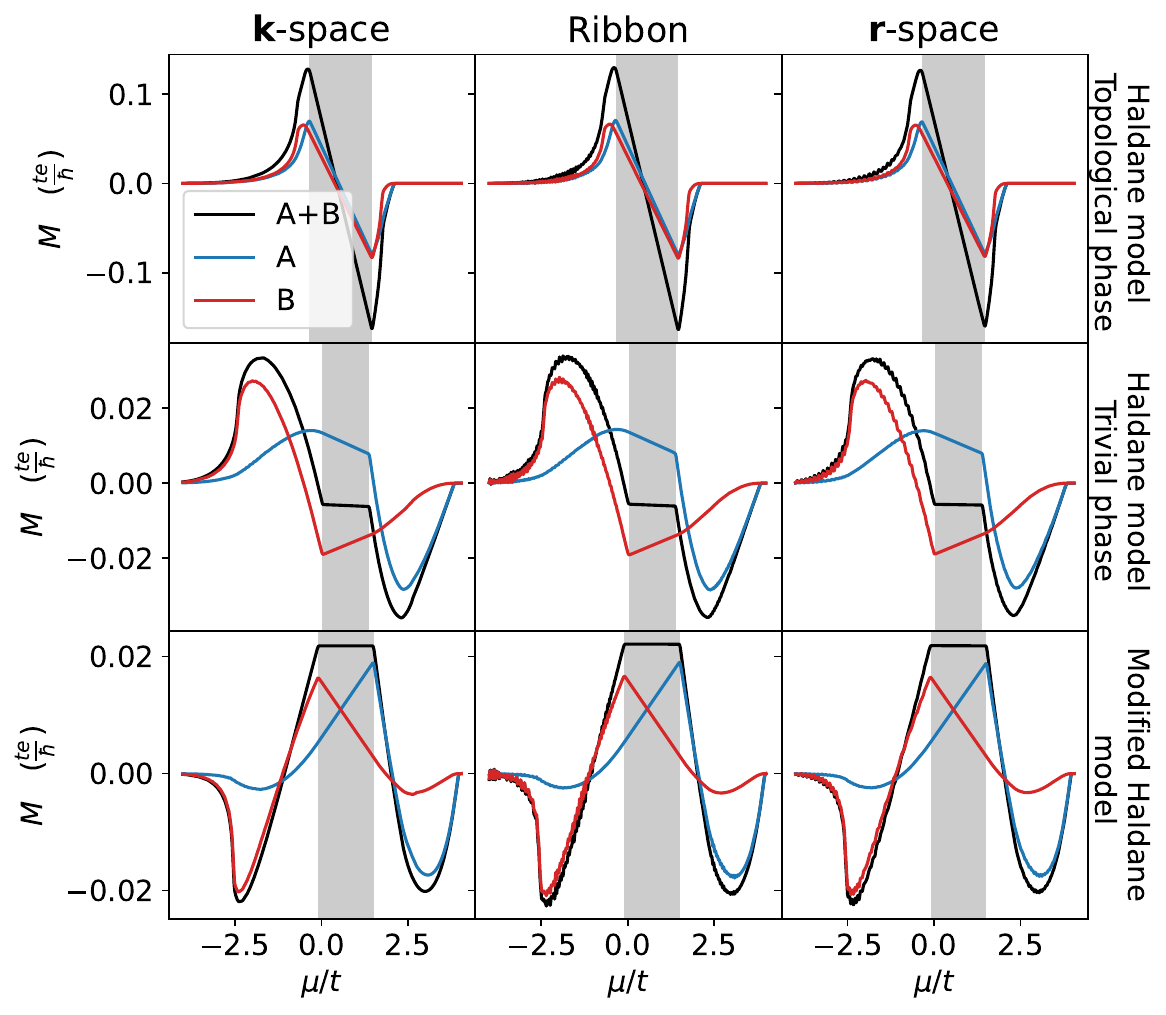}
        \caption{Site-resolved local orbital magnetizations $M_A$ (red) and $M_B$ (blue); $M_A+M_B$ (black). In gray, we highlight the gap region. We observe in the gap that the slope of the black curve is quantized and satisfies exactly $\partial_\mu(M_A+M_B) = \frac{e}{h}C$, even locally in the ribbon geometry and in $\bm r$-space. Rows: (top) HM$_{\text{topo}}$, (middle) HM$_{\text{triv}}$,(bottom)  mHM. Columns: (left) results for Bloch electron using \eqref{eq:MorbExtendedLocal}, (middle) results for ribbon geometry (see \cite{supplementary}), (right) results obtained using \eqref{eq:localMorbFin}, computed locally on a single site in the bulk, for a finite sample of $N=1088$ sites and open boundaries. $t$ denotes the nearest-neighbor hopping term on the honeycomb lattice.}
        \label{fig:sublatticeMorb}
    \end{center}
\end{figure}

{\color{blue}\textit{Application on two-band models}-}

To illustrate the above formalism, we consider three representative two-band models: the Haldane model \cite{haldane_model_1988} in its topological (HM${\rm topo}$) and trivial (HM${\rm triv}$) phases, and the modified Haldane model (mHM) \cite{Coloms2018}, which is also topologically trivial. In the mHM, the current-loop order is identical on the two triangular Bravais sublattices, whereas in the Haldane model the loop currents are related by time reversal (see Ref. \cite{supplementary} for the corresponding Bloch and real-space Hamiltonians).

Figure \ref{fig:sublatticeMorb} shows the sublattice orbital magnetization as a function of chemical potential for the three models, evaluated in the three geometries considered above. Despite the fundamentally different formulations underlying each calculation, the results are in perfect numerical agreement. This demonstrates the internal consistency of the local theory and establishes that the orbital magnetic texture is an intrinsic bulk property, independent of the chosen geometry.

The local theory further resolves qualitatively distinct orbital magnetic orders. In HM${\rm topo}$ [Fig. \ref{fig:sublatticeMorb} upper row], the two sublattices carry nearly identical orbital magnetizations, $M_A(\mu)\simeq M_B(\mu)$, indicative of orbital ferromagnetism. In contrast, HM${\rm triv}$ [Fig. \ref{fig:sublatticeMorb} middle row] exhibits orbital ferrimagnetism within the bands and nearly perfect antiferromagnetism inside the gap, becoming exactly antiferromagnetic in the particle-hole symmetric limit. The modified Haldane model [Fig. \ref{fig:sublatticeMorb} lower row] instead displays orbital ferrimagnetism both inside the bands and throughout the gap.
A particularly striking prediction of the local theory concerns the slope of orbital magnetization inside an insulating gap. This slope is entirely governed by the local Berry curvature and contains two distinct contributions. For two band systems, the first is a sublattice-independent {\em topological} contribution, equal to one half the Chern number of the filled band, while the second is a sublattice-dependent {\em geometrical} contribution, originating solely from $\Omega^{\rm geom}_{n\mathbf{k}}(\bm r)$. The latter takes opposite values on the two sublattices, vanishes for particle-hole symmetric systems, and is significant in trivial gaps [Fig. \ref{fig:sublatticeMorb}].
More generally, these results demonstrate that local orbital magnetization encodes microscopic information beyond bulk topology, highlighting the importance of atomic scale resolution in materials with complex unit cells, such as Kagome lattices, moiré systems, and three-dimensional quantum materials.

{\color{blue}\textit{Comparison with the Bianco-Resta approach}-}

Shortly after the development of the modern theory of orbital magnetization, Bianco and Resta introduced a real-space representation of orbital magnetization in finite-size systems with open boundaries at zero temperature, given by \cite{Bianco2011,Bianco2013}. 
\begin{align}
      {M}(\bm r) =\dfrac{e}{\hbar} ( \mathfrak{M}(\bm r) +  \dfrac{\mu}{2\pi} 
    \mathcal{C}(\bm r)) 
    \label{eq:LocMorbModern}
\end{align} with 
\begin{align}
    \mathfrak{M}_{\text{BR}}(\bm r) &=  \mathrm{Im} \langle \bm r| \mathcal{P} x\mathcal{Q} h\mathcal{Q} y \mathcal{P}-\mathcal{Q}x\mathcal{P} h \mathcal{P} y \mathcal{Q}  | \bm r \rangle \\
    \mathcal{C}_{\text{BR}}(\bm r) &= 4\pi
    %4\pi 
    \mathrm{Im}  \langle \bm r|\mathcal{Q}x\mathcal{P}y \mathcal{Q}  | \bm r \rangle 
\end{align} where $\mathcal{P} = \sum_{n}f(\varepsilon_n)|n\rangle\langle n|$ is the projector onto occupied states at zero temperature and $\mathcal{Q} = 1 - \mathcal{P}$ its orthogonal. 
$\mathfrak{M}_{\text{BR}}(\bm r)$ is a magnetization density (in units of $e/\hbar$) defined such that its sum over the entire system yields back the total orbital magnetization of finite systems of Eq. \eqref{eq:MorbTotrspace}.
$\mathcal{C}_{\text{BR}}(\bm r)$ is a local Chern number first introduced in \cite{Bianco2011}, satisfying $\sum_{\bm r} \mathcal{C}_{\text{BR}}(\bm r)=0$, and defined such that locally in a bulk unit-cell, its value in a topological gap is equal to the Chern number of the bulk material\cite{Seleznev2023,Drigo2020}. 
\begin{figure}[ht!]
    \begin{center}
        \includegraphics[width=0.49\textwidth]{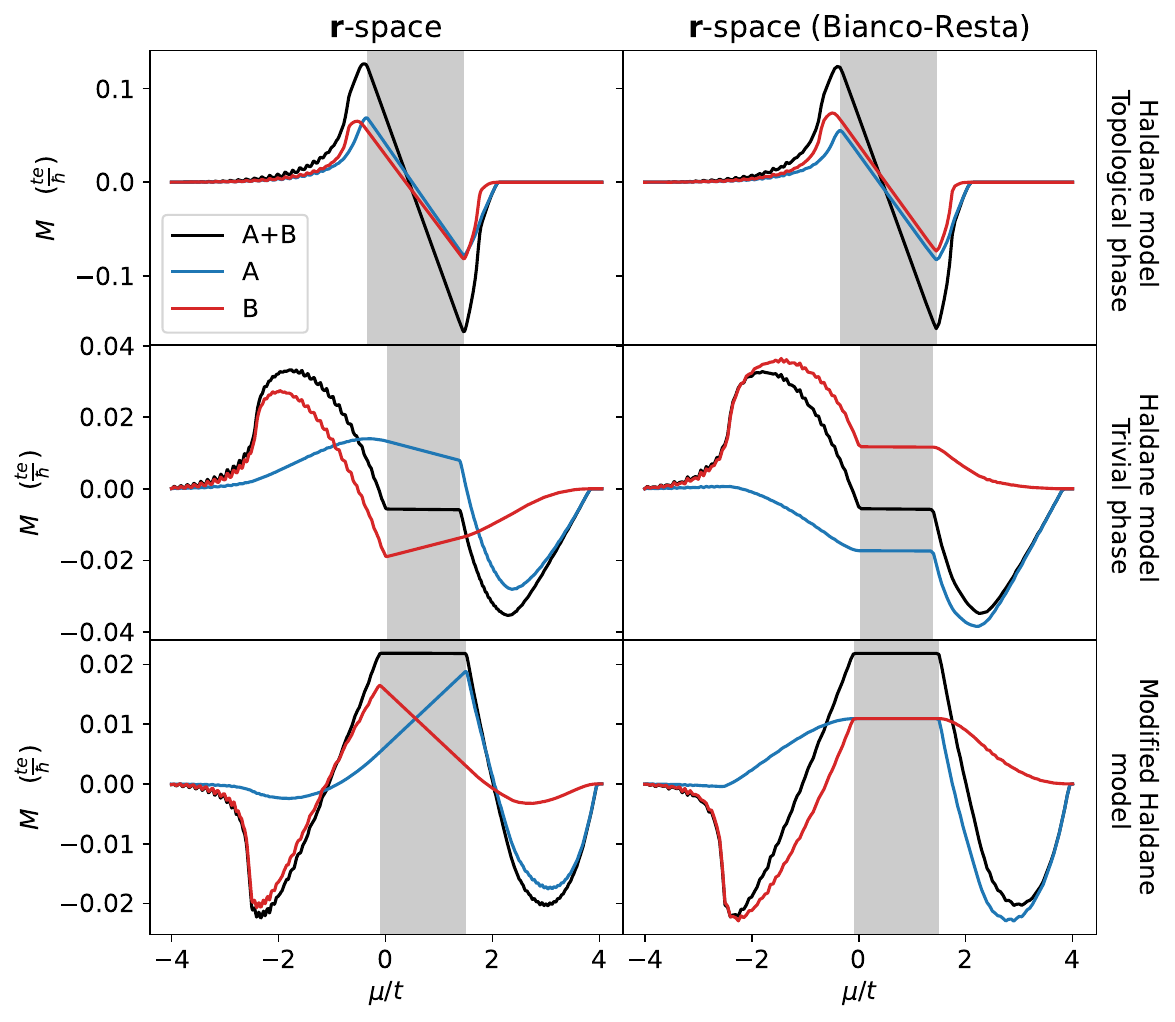}
        \caption{Magnetization for a single site (chosen arbitrarily in the bulk - all bulk sites are found to give equivalent results) in each sublattice in a sample with $N=1088$ sites and open boundaries: (Left) computed using Eqs. \eqref{eq:Localmag} and \eqref{eq:LocalChern}, (right) computed using Bianco-Resta expressions. (top) HM$_{\text{topo}}$, (middle) HM$_{\text{triv}}$, (bottom) mHM.}
        \label{fig:ComptThermoModern}
    \end{center}
\end{figure} 
In the zero-temperature limit, our theory Eq. \eqref{eq:localMorbFin}, assumes a form analogous to \eqref{eq:LocMorbModern}
with 
    \begin{align}
    &\mathfrak{M}(\bm r)=\sum_n f(\varepsilon_n) \left(  m_{n} (\bm r)- \varepsilon_n \Omega_n(\bm r) \right) ,
     \label{eq:Localmag}
  \end{align} and an associated local Chern marker
\begin{align}   
    &{\cal C }(\bm r) = 2\pi \sum_{n} f(\varepsilon_n) \Omega_n(\bm r).
    \label{eq:LocalChern}
\end{align} The present expression can be evaluated with a computational cost scaling as $N^2$ instead of the $N^3$ cost required in the Bianco-Resta formulation.
Although the two formulations share the same formal structure and recover the same bulk orbital magnetization, they are conceptually different. The Bianco-Resta approach starts from the global orbital magnetization and therefore the magnetization density $ \mathfrak{M}(\bm r)$ contains a gauge ambiguity associated with surface magnetization related to hinge currents, as discussed in \cite{Seleznev2023}.
By contrast, our formalism based on the field dependent local density of states provides a unique definition of $\mathfrak{M}(\bm r)$, without gauge ambiguity, and constitutes a transparent real-space counterpart of zero-temperature expression of the modern theory \cite{Thonhauser2005,Ceresoli2006}.

As illustrated in Fig. \ref{fig:ComptThermoModern}, the two formulations yield markedly different sublattice orbital magnetic textures, both in bands and gaps. In the gaps, the Bianco-Resta approach does not account for the slope of the magnetization that is related to the geometric part of the local Berry curvature, which is especially visible in the trivial phases.

{\color{blue}{\textit{Anatomy of bulk topology-}}} 

\begin{figure}[ht!]
    \begin{center}
    \includegraphics[width=0.49\textwidth]{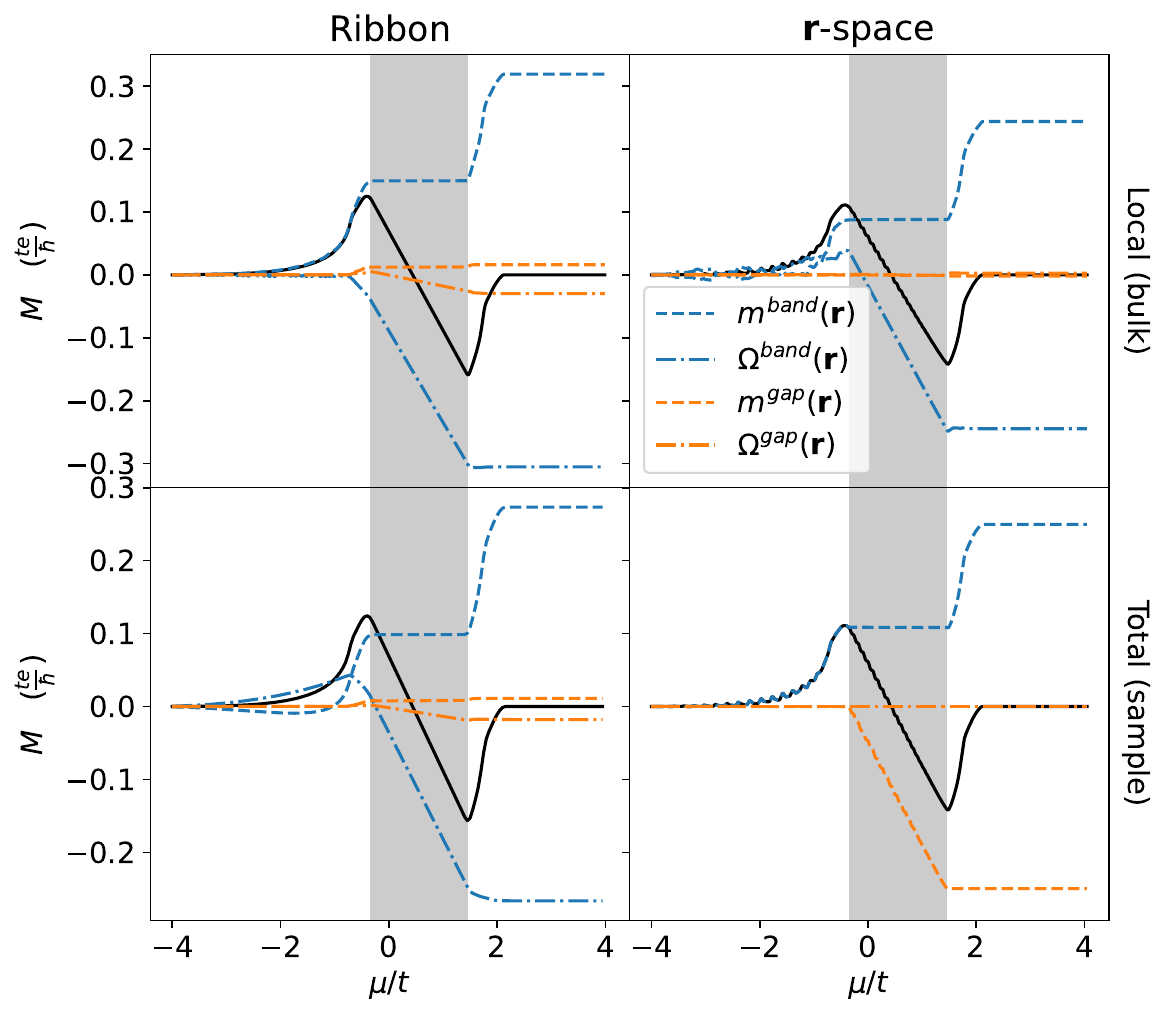}
        \caption{Orbital magnetization in the topological phase of the Haldane model  (left) in the ribbon geometry and (right) in open-boundary finite system, for (top) a single site in the bulk  and (bottom) averaged over the entire system. Blue curves: pure band states contribution. Orange curves: pure gap states and mixed band-gap states contributions. Dashed-line: orbital momentum contribution, dot-dashed: Berry curvature contribution.}
        \label{fig:ComptBandGap}
    \end{center}
\end{figure}

A remarkable consequence of the present formulation is the possibility to resolve the respective contributions of band and gap states in the local orbital magnetization.
As shown in Fig. \ref{fig:ComptBandGap}, at the local level, both in ribbons and finite flakes, the topological response is found to originate entirely from the local Berry-curvature contribution of band states. Gap states do not contribute to the local bulk topological response. By contrast, after averaging over the entire sample, the same quantized slope of orbital magnetization is recovered through different microscopic mechanisms depending on the geometry. In ribbons, it is carried by the Berry curvature of band states, while, in finite systems with open boundaries, the Berry-curvature contribution vanishes upon summation over the sample, and the quantized slope is instead recovered through the orbital moments of gap states.  This decomposition clarifies how distinct microscopic mechanisms cooperate to produce the same quantized macroscopic response and makes explicit, within a unified thermodynamic framework, earlier observations that local bulk topology can be identified independently of edge states \cite{Marrazzo2016,Bianco2016}.

{\color{blue}{\textit{Conclusion-}}} 

We have developed a microscopic thermodynamic theory of local orbital magnetization, hence providing a natural generalization of the modern theory from a unit-cell averaged quantity to the atomic scale. Starting from the magnetic-field dependence of the local density of states, we derive a local orbital magnetization that naturally separates into a local orbital-moment contribution, describing the magnetic-field-induced shift of the electronic spectrum, and a previously unidentified local Berry curvature describing the field-induced redistribution of electronic weight in real space.

The theory is valid in periodic crystals, ribbon geometries, and finite systems with open boundaries, providing a unified microscopic description of orbital magnetization across all geometries while recovering the modern theory after averaging over a unit cell. Unlike previous real-space formulations, which construct a local representation of orbital magnetization starting from the total magnetization, the present approach derives the microscopic local magnetic response itself. The associated local Berry curvature naturally defines a local Chern marker while revealing an additional geometric contribution that is completely hidden in unit-cell-averaged formulations.

Beyond resolving orbital magnetic textures at the single-site scale, the present theory provides a microscopic picture of how topology is encoded in real space. It reveals that the locally in the bulk, the topological response is entirely carried by the Berry curvature of band states, while the macroscopic orbital magnetization may emerge through different microscopic mechanisms depending on the geometry, thereby offering a new perspective on bulk-edge correspondence.

More broadly, the present framework establishes local orbital magnetization as a genuine microscopic observable. It opens the way to the study of orbital magnetic textures and local thermodynamic responses in systems lacking translational symmetry (including molecules, amorphous materials, quasicrystals, defects, and moiré structures) \cite{Lage2026} as well as to the development of microscopic theories of higher-order responses such as orbital susceptibility, of other observables such as local spin magnetization \cite{SticletPiechon2026}, or for interacting systems \cite{Acheche2019,Kang2026,Chen2026,PeraltaGavensky2023,Ye2026,Huang2026}.

{\color{blue}{\it Acknowledgment:}} F.P acknowledges useful discussions with Jean-Noël Fuchs, Yasuhiro Hatsugai, Andrej Mesaros, and Tomonari Mizoguchi.

\bibliographystyle{apsrev4-1}
\bibliography{biblioPaper}

@article{haldane_model_1988,
  title      = {Model for a {Quantum} {Hall} {Effect} without {Landau} {Levels}: {Condensed}-{Matter} {Realization} of the "{Parity} {Anomaly}"},
  volume     = {61},
  issn       = {0031-9007},
  shorttitle = {Model for a {Quantum} {Hall} {Effect} without {Landau} {Levels}},
  doi        = {10.1103/PhysRevLett.61.2015},
  language   = {en},
  number     = {18},
  journal    = {Physical Review Letters},
  author     = {Haldane, F. D. M.},
  month      = oct,
  year       = {1988},
  pages      = {2015--2018}
}

@article{Berry,
  title    = {Quantal phase factors accompanying adiabatic changes},
  volume   = {392},
  issn     = {},
  language = {en},
  number   = {},
  journal  = {Proceedings of the Royal Society A, Mathematical Physical and Engineering Sciences},
  author   = {Berry, M. V.},
  month    = sep,
  year     = {1984},
  pages    = {1802}
}

@article{Bianco2013,
  title = {Orbital Magnetization as a Local Property},
  volume = {110},
  ISSN = {1079-7114},
  DOI = {10.1103/physrevlett.110.087202},
  number = {8},
  journal = {Physical Review Letters},
  publisher = {American Physical Society (APS)},
  author = {Bianco,  Raffaello and Resta,  Raffaele},
  year = {2013},
  month = feb 
}

@article{Raoux2015,
  title = {Orbital magnetism in coupled-bands models},
  volume = {91},
  ISSN = {1550-235X},
  DOI = {10.1103/physrevb.91.085120},
  number = {8},
  journal = {Physical Review B},
  publisher = {American Physical Society (APS)},
  author = {Raoux,  Arnaud and Piéchon,  Frédéric and Fuchs,  Jean-Noël and Montambaux,  Gilles},
  year = {2015},
  month = feb 
}

@article{Thonhauser2011,
  title = {THEORY OF ORBITAL MAGNETIZATION IN SOLIDS},
  volume = {25},
  ISSN = {1793-6578},
  DOI = {10.1142/s0217979211058912},
  number = {11},
  journal = {International Journal of Modern Physics B},
  publisher = {World Scientific Pub Co Pte Lt},
  author = {Thonhauser,  T.},
  year = {2011},
  month = apr,
  pages = {1429–1458}
}

@article{Thonhauser2005,
  title = {Orbital Magnetization in Periodic Insulators},
  volume = {95},
  ISSN = {1079-7114},
  DOI = {10.1103/physrevlett.95.137205},
  number = {13},
  journal = {Physical Review Letters},
  publisher = {American Physical Society (APS)},
  author = {Thonhauser,  T. and Ceresoli,  Davide and Vanderbilt,  David and Resta,  R.},
  year = {2005},
  month = sep 
}

@article{Ceresoli2006,
  title = {Orbital magnetization in crystalline solids: Multi-band insulators,  Chern insulators,  and metals},
  volume = {74},
  ISSN = {1550-235X},
  DOI = {10.1103/physrevb.74.024408},
  number = {2},
  journal = {Physical Review B},
  publisher = {American Physical Society (APS)},
  author = {Ceresoli,  Davide and Thonhauser,  T. and Vanderbilt,  David and Resta,  R.},
  year = {2006},
  month = jul 
}

@article{Xiao2005,
  title = {Berry Phase Correction to Electron Density of States in Solids},
  volume = {95},
  ISSN = {1079-7114},
  DOI = {10.1103/physrevlett.95.137204},
  number = {13},
  journal = {Physical Review Letters},
  publisher = {American Physical Society (APS)},
  author = {Xiao,  Di and Shi,  Junren and Niu,  Qian},
  year = {2005},
  month = sep 
}

@article{Peierls1933,
  title = {Zur Theorie des Diamagnetismus von Leitungselektronen},
  volume = {80},
  ISSN = {1434-601X},
  DOI = {10.1007/bf01342591},
  number = {11–12},
  journal = {Zeitschrift fur Physik},
  publisher = {Springer Science and Business Media LLC},
  author = {Peierls,  R.},
  year = {1933},
  month = nov,
  pages = {763–791}
}

@article{Streda1982,
  title = {Theory of quantised Hall conductivity in two dimensions},
  volume = {15},
  ISSN = {0022-3719},
  DOI = {10.1088/0022-3719/15/22/005},
  number = {22},
  journal = {Journal of Physics C: Solid State Physics},
  publisher = {IOP Publishing},
  author = {Streda,  P},
  year = {1982},
  month = aug,
  pages = {L717–L721}
}

@article{Bianco2011,
  title = {Mapping topological order in coordinate space},
  volume = {84},
  ISSN = {1550-235X},
  DOI = {10.1103/physrevb.84.241106},
  number = {24},
  journal = {Physical Review B},
  publisher = {American Physical Society (APS)},
  author = {Bianco,  Raffaello and Resta,  Raffaele},
  year = {2011},
  month = dec 
}

@article{Bianco2016,
  title = {Orbital magnetization in insulators: Bulk versus surface},
  volume = {93},
  ISSN = {2469-9969},
  DOI = {10.1103/physrevb.93.174417},
  number = {17},
  journal = {Physical Review B},
  publisher = {American Physical Society (APS)},
  author = {Bianco,  Raffaello and Resta,  Raffaele},
  year = {2016},
  month = may 
}

@misc{supplementary,
  title = {Supplementary Material - {T}heory of local orbital magnetization: {L}ocal Berry curvature},
  author = {Al Saati, Sariah and le Hur, Karyn and  Piéchon, Frédéric},
  year = {2025}
}

@article{Xiao2010,
  title = {Berry phase effects on electronic properties},
  volume = {82},
  ISSN = {1539-0756},
  DOI = {10.1103/revmodphys.82.1959},
  number = {3},
  journal = {Reviews of Modern Physics},
  publisher = {American Physical Society (APS)},
  author = {Xiao,  Di and Chang,  Ming-Che and Niu,  Qian},
  year = {2010},
  month = jul,
  pages = {1959–2007}
}

@article{Marrazzo2016,
  title = {Irrelevance of the Boundary on the Magnetization of Metals},
  volume = {116},
  ISSN = {1079-7114},
  DOI = {10.1103/physrevlett.116.137201},
  number = {13},
  journal = {Physical Review Letters},
  publisher = {American Physical Society (APS)},
  author = {Marrazzo,  Antimo and Resta,  Raffaele},
  year = {2016},
  month = apr 
}

@article{Drigo2020,
  title = {Chern number and orbital magnetization in ribbons,  polymers,  and single-layer materials},
  volume = {101},
  ISSN = {2469-9969},
  DOI = {10.1103/physrevb.101.165120},
  number = {16},
  journal = {Physical Review B},
  publisher = {American Physical Society (APS)},
  author = {Drigo,  Enrico and Resta,  Raffaele},
  year = {2020},
  month = apr 
}

@article{Shi2007,
  title = {Quantum Theory of Orbital Magnetization and Its Generalization to Interacting Systems},
  volume = {99},
  ISSN = {1079-7114},
  DOI = {10.1103/physrevlett.99.197202},
  number = {19},
  journal = {Physical Review Letters},
  publisher = {American Physical Society (APS)},
  author = {Shi,  Junren and Vignale,  G. and Xiao,  Di and Niu,  Qian},
  year = {2007},
  month = Nov 
}

@article{Seleznev2023,
  title = {Towards a theory of surface orbital magnetization},
  volume = {107},
  ISSN = {2469-9969},
  DOI = {10.1103/physrevb.107.115102},
  number = {11},
  journal = {Physical Review B},
  publisher = {American Physical Society (APS)},
  author = {Seleznev,  Daniel and Vanderbilt,  David},
  year = {2023},
  month = mar 
}

@article{Kang2026,
  author = {Kang,  Jian and Wang,  Minxuan and Vafek,  Oskar},
  title = {Orbital magnetization and magnetic susceptibility of interacting electrons},
  journal = {arXiv2509.20626},
  year = {2025}
}

@article{Chen2026,
  author = {Chen,  Xi and Song,  Zhi-Da},
  title = {Orbital Magnetization of Interacting Electrons},
  journal = {arXiv2602.02478},
  year = {2026}
}

@article{PeraltaGavensky2023,
  title = {Connecting the Many-Body Chern Number to Luttinger’s Theorem through Středa’s Formula},
  volume = {131},
  ISSN = {1079-7114},
  DOI = {10.1103/physrevlett.131.236601},
  number = {23},
  journal = {Physical Review Letters},
  publisher = {American Physical Society (APS)},
  author = {Peralta Gavensky,  Lucila and Sachdev,  Subir and Goldman,  Nathan},
  year = {2023},
  month = dec 
}

@article{Ye2026,
  author = {Ye,  Mengxing},
  title = {A Quantum Many-Body Approach for Orbital Magnetism in Correlated Multiband Electron Systems},
  journal = {arXiv2601.14372},
  year = {2026}
}

@article{SticletPiechon2026,
  author = {Sticlet,  Doru and Piéchon,  Frédéric},
  title = {Local spin magnetization in itinerant non-collinear magnets: The local spin Berry curvature},
  journal = {arXiv2607.06253},
  year = {2026}
}

@article{Lage2026,
  title = {Orbital magnetization in Sierpinski fractals},
  volume = {113},
  ISSN = {2469-9969},
  DOI = {10.1103/wshz-mrnb},
  number = {12},
  journal = {Physical Review B},
  publisher = {American Physical Society (APS)},
  author = {Lage,  L. L. and Cysne,  Tarik P. and Latgé,  A.},
  year = {2026},
  month = Mar 
}

@article{Huang2026,
  author = {Huang,  Chunli},
  title = {Orbital Magnetization from Uniform and Periodic Magnetic Fields},
  journal = {arXiv2605.26889},
  year = {2026}
}

@article{Acheche2019,
  title = {Orbital magnetization and anomalous Hall effect in interacting Weyl semimetals},
  volume = {99},
  ISSN = {2469-9969},
  DOI = {10.1103/physrevb.99.075144},
  number = {7},
  journal = {Physical Review B},
  publisher = {American Physical Society (APS)},
  author = {Acheche,  S. and Nourafkan,  R. and Tremblay,  A.-M. S.},
  year = {2019},
  month = Feb 
}

@article{Coloms2018,
  title = {Antichiral Edge States in a Modified Haldane Nanoribbon},
  volume = {120},
  ISSN = {1079-7114},
  DOI = {10.1103/physrevlett.120.086603},
  number = {8},
  journal = {Physical Review Letters},
  publisher = {American Physical Society (APS)},
  author = {Colomés,  E. and Franz,  M.},
  year = {2018},
  month = Feb 
}
\end{document}

\nobibliography*
\begin{center}
    \LARGE{Supplementary Material - Theory of local orbital magnetization: local Berry curvature}

    \large{Sariah Al Saati$^1$ , Karyn le Hur$^1$, Frédéric Piéchon$^2$ - July 19, 2026}

    {\small $^1$ Centre de Physique Théorique, CNRS, École polytechnique, IPP, 91120 Palaiseau, France \\
    $^2$ Laboratoire de Physique des Solides, Université Paris-Saclay, CNRS, 91405 Orsay, France}
\end{center}

In this supplementary material, we introduce in the first section a generalized Haldane that provides the three cases discussed in the main text. In the second section, we review the perturbative computation of the first order field-dependent correction to the local density of states, that allows us to obtain a definition of the local orbital magnetization. In the third section, we detail how this definition can be used to obtain explicit expressions of local orbital magnetization for all relevant geometries. The main equations of this document are Eq. \eqref{eq:MorbGreenArticle} providing the Green's function definition of local orbital magnetization, Eq. \eqref{eq:MorbFinsize} giving the expression of local orbital magnetization in finite systems with open boundaries, Eq. \eqref{eq:Morbkspace} for periodic systems, and Eq. \eqref{eq:Morbribbon} for the ribbon geometry.

\tableofcontents

\section{Generalized Haldane model}
\label{sec:Models}

\begin{figure}[!ht]
    \begin{center}
        \includegraphics[width=0.45\textwidth]{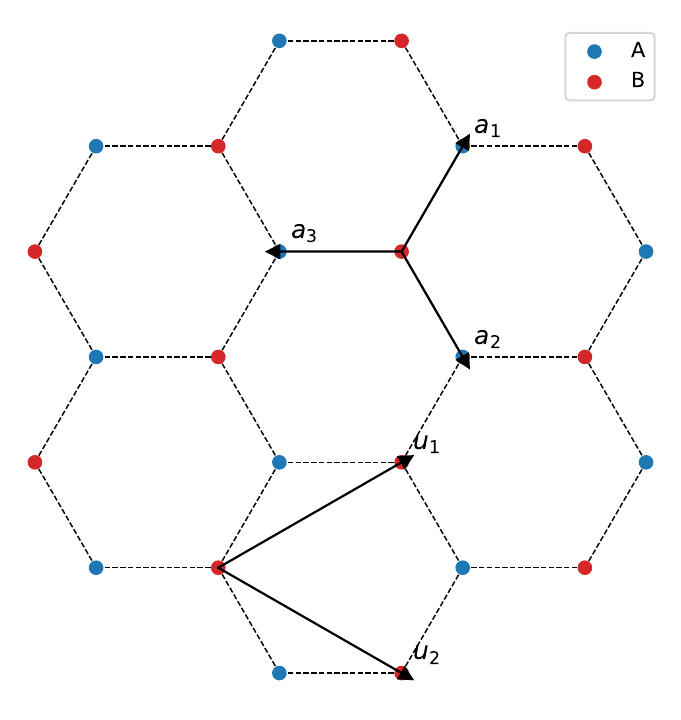}
        \includegraphics[width=0.4\textwidth]{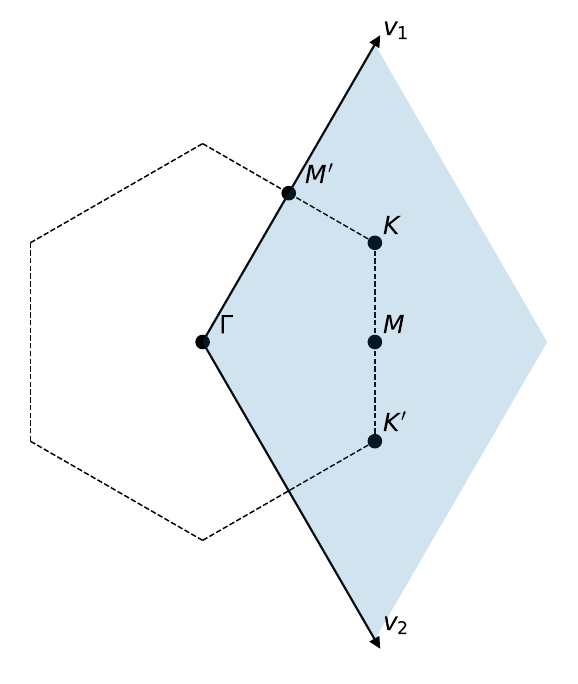}
    \end{center}
    \caption{(Left) Honeycomb lattice described as a lattice of two non-equivalent sites per unit-cell span by the two lattice vectors $\bm u_1$ and $\bm u_2$. (Right) Hexagonal shape of the Brillouin zone of the honeycomb lattice along with its equivalent description as the losange span by the reciprocal lattice vectors $\bm v_1$ and $\bm v_2$ shown as a shaded area. The second definition is the one used throughout the manuscript for the numerical computations.}
    \label{fig:HC}
\end{figure}

The generalized Haldane model \cite{haldane_model_1988} is  a tight-binding model of spinless electrons in a honeycomb lattice as shown in figure \ref{fig:HC} (left) and is span by two primitive lattice vectors $\bm{u}_1$, $\bm{u}_2$. \begin{align}
    \bm{u}_1 & = \dfrac{a}{2} \left(3, \sqrt{3} \right)  &
    & \bm{u}_2  = \dfrac{a}{2} \left(3, -\sqrt{3} \right)
\end{align} The unit cell of the lattice contains two sites, which we denote as A (blue) and B (red). This lattice is described by the nearest-neighbor vectors $\bm{a}_1$, $\bm{a}_2$, $\bm{a}_3$ and the next-nearest-neighbor vectors $\bm{b}_1$, $\bm{b}_2$, $\bm{b}_3$.  These vectors are given by 
\begin{align}
    \bm{a}_1 &=  \dfrac{a}{2} \left(1, \sqrt{3} \right) & &\bm{b}_1 = \dfrac{a}{2} \left(-3, \sqrt{3} \right) \\
    \bm{a}_2 &=  \dfrac{a}{2} \left(1, -\sqrt{3} \right) &  &\bm{b}_2 = \dfrac{a}{2} \left(3, \sqrt{3} \right)\\
    \bm{a}_3 &=   a\left(-1, 0 \right) & &\bm{b}_3 = a\left(0, -\sqrt{3} \right)
\end{align} 
where $a=1$ is the distance between two nearest-neighbors. The reciprocal lattice of this system also has a honeycomb structure. Its first Brillouin zone is described in figure \ref{fig:HC} (right), where the main high-symmetry points are described. This reciprocal lattice is span by the two reciprocal primitive vectors $\bm{v}_1$, $\bm{v}_2$ with:
\begin{align}
    \bm{v}_1 & = \dfrac{2\pi}{3a} \left(1, \sqrt{3} \right)  &
    & \bm{v}_2  = \dfrac{2\pi}{3a} \left(1, -\sqrt{3} \right)
\end{align}
The main high-symmetry points of the Brillouin zone are given by:  \begin{align}
    \label{eq:pK}
    \bm{K} & = \dfrac{2\pi}{3a} \left(1, \dfrac{1}{\sqrt{3}}\right)  & \bm{K}^\prime & = \dfrac{2\pi}{3a} \left(1, -\dfrac{1}{\sqrt{3}}\right)  \\
    \bm{M} & = \dfrac{2\pi}{3a} \left(1, 0\right)  & \bm{M}^\prime & = \dfrac{2\pi}{3a} \left(1, \sqrt{3}\right)  
    \label{eq:pM}
\end{align}
In the second quantification, we denote  $c_{\bm{r}}$ the annihilation operator of a spinless electron at site $\bm r$; satisfying the usual anti-commutation rule \begin{align}
    \left\{c_{\bm{r}}, {c^\dagger}_{\bm{r}^\prime}\right\} = \delta_{\bm{r}, \bm{r}^\prime}
\end{align} For more convenience, we introduce the following operators 
\begin{align}
    c_{A, \bm{r}} &= {c}_{\bm{r} + \bm{a}_3} \\
    c_{B, \bm{r}} &= {c}_{\bm{r}} 
\end{align}
where now the position variable $\bm{r}$ describes the Bravais lattice position of the considered unit-cell. 
In this real space representation , the generalized Haldane model is given by the following Hamiltonian \begin{align}
    \mathcal{H} &=  \mathcal{H}_1 +  \mathcal{H}_M +  \mathcal{H}_2
\end{align}
where $\mathcal{H}_{1,2}$ and $\mathcal{H}_M$ describe different hopping processes. 
The first term $\mathcal{H}_1$ describes the nearest-neighbor hopping processes of amplitude $t=1$ such that 
\begin{align}
    \mathcal{H}_1 &= - t \sum_{\bm{r}} \sum_{\bm{\alpha}} \left[ {c^\dagger}_{B, \bm{r}} {c}_{A, \bm{r} + \bm{\alpha}} + {c^\dagger}_{A, \bm{r} + \bm{\alpha}}{c}_{B, \bm{r}}\right] 
    \label{eq:H1}
\end{align}
where the sum over $\bm{\alpha}$ runs over the three vectors $\bm{0}$, $\bm{u}_1 $ and $\bm{u}_2$. The next term $\mathcal{H}_M$ describes an onsite potential that breaks inversion symmetry:
\begin{align}
    \mathcal{H}_M = + M \sum_{\bm{r}} \left[ {c^\dagger}_{A, \bm{r}}{c}_{A, \bm{r}} - {c^\dagger}_{B, \bm{r}}{c}_{B, \bm{r}}\right]
    \label{eq:HM}
\end{align}

\begin{figure}
     \centering
     \begin{subfigure}[b]{0.49\textwidth}
         \centering
         \includegraphics[trim={150 150 150 50}, clip, width=.5\textwidth]{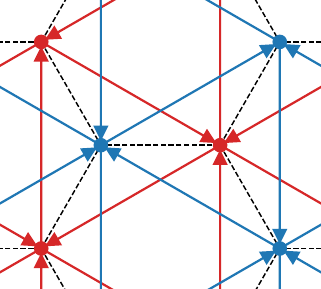}
         \caption{Haldane model \cite{haldane_model_1988}}
     \end{subfigure}
     \hfill
     \begin{subfigure}[b]{0.49\textwidth}
         \centering
         \includegraphics[trim={150 150 150 50}, clip, width=.5\textwidth]{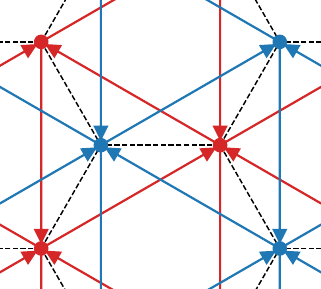}
         \caption{Modified Haldane model \cite{Coloms2018}}
     \end{subfigure}
        \caption{Next-nearest neighbor hopping term for the Haldane model \cite{haldane_model_1988} (left) and the modified Haldane model \cite{Coloms2018} (right), with the arrow showing the direction in which the fluxes are accumulated.}
    \label{fig:T2}
\end{figure}

The last term $\mathcal{H}_2$ is a generalized version of the term introduced by Haldane \cite{haldane_model_1988} in 1988. It describes next-nearest-neighbor hopping processes of amplitude $t_2$, and magnetic fluxes $\phi_{A,B}$ oriented as in figure \ref{fig:T2} :
\begin{align}
    \mathcal{H}_2 &= t_2 \sum_{\bm{r} } \sum_{i=1}^3 e^{i\phi_A}{c^\dagger}_{A, \bm{r}} {c}_{A, \bm{r}+\bm{b}_i} + e^{i\phi_B} {c^\dagger}_{B, \bm{r}} {c}_{B, \bm{r}+\bm{b}_i}  + hc
\end{align} 
This term breaks time-reversal symmetry and creates a loop current order on each sublattice.
This model gives back the original Haldane model whenever $\phi_A = -\phi_B = \phi$ 
(e.g time reversed loop current order on each sublattice). 
Applying Bloch's theorem, we obtain in reciprocal space  \begin{align}
    \mathcal{H}(\bm k) = \varepsilon_0(\bm k) + \bm d (\bm k) \cdot \bm \sigma 
    \label{eq:hamiltonian2band}
\end{align} where the Pauli matrices $\bm \sigma$ act of the $A/B$ degrees of freedom. 
We have \begin{align}
    \varepsilon_0 (\bm k) &=  t_2 (\cos \phi_A + \cos \phi_B) \sum_{j=1}^3  \cos\left(\bm{k}\cdot \bm{b}_j\right) - t_2 (\sin \phi_A + \sin \phi_B) \sum_{j=1}^3  \sin\left(\bm{k}\cdot \bm{b}_j\right)  \\
    d_x (\bm k) &+i d_y (\bm k) = - t \sum_{j=1}^3 e^{i\bm{k}\cdot \bm{a}_j}  \\
    d_z (\bm k) &=   M + t_2 (\cos \phi_A - \cos \phi_B) \sum_{j=1}^3  \cos\left(\bm{k}\cdot \bm{b}_j\right) - t_2 (\sin \phi_A - \sin \phi_B) \sum_{j=1}^3  \sin\left(\bm{k}\cdot \bm{b}_j\right) 
\end{align} At the Dirac points,  we have \begin{align}
    d_z(\bm K_\xi) = M - 3t_2\dfrac{\cos \phi_A - \cos \phi_B}{2} + 3\sqrt{3} t_2 \xi \dfrac{\sin \phi_A - \sin \phi_B}{2} 
\end{align} 

For this generalized Haldane model, the necessary condition to be in a topological phase with a finite Chern number is given by:
\begin{align}
    \left| M -  3t_2\dfrac{\cos \phi_A - \cos \phi_B}{2}\right| < \left|3\sqrt{3} t_2  \dfrac{\sin \phi_A - \sin \phi_B}{2} \right|.
\end{align}

In the main text, we consider three realizations of this model: \begin{itemize}
    \item the Haldane model in a topological phase ($HM_\text{topo}$), characterized by the parameters $\phi_A = -\phi_B = \pi/4$, $t_2 = 1/3$, $M=\sqrt{3}/10$ in energy units of $t =1$. 
    \item the Haldane model in a trivial phase ($HM_\text{triv}$), characterized by the parameters $\phi_A = -\phi_B = \pi/4$, $t_2 = 1/3$, $M=1.1\sqrt{3}$ in energy units of $t =1$.
    \item the modified Haldane model ($mHM$) \cite{Coloms2018}, necessarily trivial, characterized by the parameters  $\phi_A = \phi_B = \pi/4$, $t_2 = 1/3$, $M=1.17\sqrt{3}$ in energy units of $t =1$. 
\end{itemize}

\section{Perturbation theory}

In a thermodynamic picture, the spontaneous local orbital magnetization measures the first-order correction of the local density of states with respect to a small magnetic field. For a non-interacting system in a single-particle picture, the grand potential in the grand-canonical ensemble is given by \begin{align}
    \Xi(\bm r, \mu, B) &= - \int_{-\infty}^{+\infty}\mathrm{d}\varepsilon F(\varepsilon) \rho(\bm r,\varepsilon, B)
\end{align} where $F(\varepsilon) = k_B T \ln\left(1 + e^{-\beta(\varepsilon-\mu)}\right)$. In this expression, $\rho(\bm r,\varepsilon, B)$ is the local density of states whose trace yields the Density of States of the system. It is expressed in terms of the diagonal terms of the Green's function of the system $\mathcal{G}(\varepsilon, B)$ as \begin{align}
    \rho(\bm r,\varepsilon, B) &= - \dfrac{1}{\pi} \mathrm{Im} \mathcal{G}_{{\bm r}{\bm r}}(\varepsilon, B)
\end{align}
In this context, the local orbital magnetization is defined as
\begin{align}
    \bm M_{\text{orb}}(\bm r) = - \left. \dfrac{\partial \Xi_{\bm r}(T, \mu, B)}{\partial \bm B} \right|_{\bm B\rightarrow \bm 0}
\end{align} 

The perturbative expansion of the Green's function at first order in $\bm B$ is $\mathcal{G}(\varepsilon,\bm B) = \mathcal{G}(\varepsilon) +  {\mathcal{G}}^{(1)}(\varepsilon,\bm B) + O(|\bm B|^2)$ with ${\mathcal{G}}^{(1)}(\varepsilon,\bm B) =\bm B \cdot \partial_{\bm B}\mathcal{G}|_{\bm B \rightarrow \bm 0}$. Correspondingly, the local orbital magnetization is given by \begin{align}
    \bm M_{\text{orb}}(\bm r) = - \dfrac{1}{\pi}\int_{-\infty}^{+\infty}\mathrm{d}\varepsilon F(\varepsilon)  \mathrm{Im}  \partial_{\bm B}\mathcal{G}^{(1)}_{{\bm r}{\bm r}}(\varepsilon,\bm B)|_{\bm B\rightarrow \bm 0} 
\end{align} In the following, we are thus interested in computing an expression of $ {\mathcal{G}}^{(1)}$ in terms of the zero-field Green's function $ \mathcal{G}(\varepsilon)$. This computation was done by \citet{Raoux2015}, and we review it in this section. In the following, we adopt the notations where indices $i,j,k$ describe site positions in position space instead of $\bm r, \bm r^\prime, \bm r^{\prime \prime}$. We also focus in the case of a 2D material with an out-of-plane external magnetic field $\bm B = B \bm e_z$, and we only compute the $z$ component of the orbital magnetization. The generalization to an arbitrary magnetic field in a 3D material is straightforward.

We consider a generic lattice system in the tight-binding approximation described by a general zero field Hamiltonian \begin{align}
    \mathcal{H} = \sum_{ij} |i\rangle t_{ij} \langle j|
\end{align} in which the state $|i\rangle$ may actually describe a multi-orbital wavefunction at site $i$, but we focus on the case where there is only one single orbital per site. Within the tight-binding approximation, \citet{Peierls1933} showed that the effects of a slowly varying external magnetic field $\bm B = \bm \nabla \times \bm A$ can be taken into account by introducing a magnetic phase associated to each hopping term, called Peierls phase, according to 
\begin{align}
    \mathcal{H}^{B} =  \sum_{ij} |i\rangle t_{ij} e^{i \varphi_{ij}} \langle j|
\end{align} where the Peierls is defined as the line-integral of the potential vector $\bm A$  from site $i$ to site $j$ \begin{align}
    \varphi_{ij} = \dfrac{e}{\hbar} \int_{\bm r_i}^{\bm r_j} \mathrm{d} \bm l \cdot \bm A
\end{align}  Peierls phases are gauge-dependant quantities, but the circulation of the gauge potential along a closed contour is a gauge-invariant quantity according to Stokes's theorem. We thus introduce the gauge-invariant magnetic flux \begin{align}
    \Phi_{ikj} &= \varphi_{ik} + \varphi_{kj} + \varphi_{ji} = \dfrac{e}{\hbar} \oint_{ikj} \mathrm{d} \bm l \cdot \bm A = \dfrac{e}{\hbar} \iint_{ikj} \bm B \cdot \mathrm{d}\bm S
\end{align} describing the magnetic flux flowing across a triangular contour described by arbitrary sites $i$, $k$ and $j$. For a uniform magnetic field, we explicitly have \begin{align}
    \Phi_{ikj} = \dfrac{eB}{2\hbar}\left[(x_i-x_k)(y_k-y_j) - (y_i - y_k)(x_k-x_j)\right].
\end{align}
The zero-field and field-dependent  Green's  functions verify
\begin{align}
    \left(\varepsilon - \mathcal{H}\right)\mathcal{G} &= I\\
    \left(\varepsilon - \mathcal{H}^B\right)\mathcal{G}^B &= I
\end{align} 
Following \citet{Raoux2015}, we introduce a modified Green's function $\widetilde {\mathcal{G}}$ such that $\widetilde {\mathcal{G}}_{ij} = \mathcal{G}_{ij}e^{i\varphi_{ij}} $ which will provide a perturbative expansion with gauge-invariant terms. This gives us \begin{align}
    \left[  \left(\varepsilon - \mathcal{H}^B\right) \widetilde {\mathcal{G}}\right]_{ij} &= \sum_k \left(\varepsilon\delta_{ik} - t_{ik} e^{i\varphi_{ik}}  \right) e^{i\varphi_{kj}}  \mathcal{G}_{kj} \\
    &=  \sum_k \left(\varepsilon e^{i\varphi_{ij}}\delta_{ik} - t_{ik} e^{i\varphi_{ik} + i\varphi_{kj}}\right)   \mathcal{G}_{kj}  \\
    &= e^{i\varphi_{ij}} \sum_k \left(\varepsilon \delta_{ik} - t_{ik} e^{i\Phi_{ikj}}\right)   \mathcal{G}_{kj}  \\
    &= e^{i\varphi_{ij}} \sum_k \left(\varepsilon \delta_{ik} - t_{ik} + t_{ik} - t_{ik} e^{i\Phi_{ikj}}\right)   \mathcal{G}_{kj} \\
    &= e^{i\varphi_{ij}} \sum_k \left(\varepsilon \delta_{ik} - t_{ik} \right)   \mathcal{G}_{kj} + \sum_k t_{ik} \left( 1 - e^{i\Phi_{ikj}}\right)   \mathcal{G}_{kj} \\
    \left[  \left(\varepsilon - \mathcal{H}^B\right) \widetilde {\mathcal{G}}\right]_{ij} &= \left[ I - \mathcal{T}(\varepsilon)\right]_{ij}
\end{align} We deduce \begin{align}
    {(\mathcal{G}^B)}^{-1} \widetilde {\mathcal{G}} = I - \mathcal{T}
\end{align} which gives us \begin{align}
    \mathcal{G}^B = \widetilde {\mathcal{G}} \sum_{n>0} \mathcal{T}^n = \widetilde {\mathcal{G}} \left(I + \mathcal{T} + \mathcal{T}^2 + o(B^2) \right)
\end{align}

We are only interested in the first-order correction to $\mathcal{G}^B$. We know from the previous equations that  $\mathcal{T} \propto B + O(B^2)$, so that the first order correction in $B$ is given by the first power of $\mathcal{T}$ in  the above expansion  \begin{align}
    \mathcal{G}^{B}_{ii} &= \mathcal{G}_{ii} +  \sum_j e^{i\varphi_{ij}} \mathcal{G}_{ij} \mathcal{T}_{ji} \\
    &=  \mathcal{G}_{ii} + \sum_j e^{i\varphi_{ij}} \mathcal{G}_{ij} e^{i\varphi_{ji}}  \sum_k t_{jk} \left(e^{i\Phi_{jki}}- 1 \right)   \mathcal{G}_{ki}  \\
    \mathcal{G}^{B}_{ii}&=  \mathcal{G}_{ii} +\sum_{jk} i\Phi_{jki}  \mathcal{G}_{ij} t_{jk}   \mathcal{G}_{ki} 
\end{align} We introduce the position operator $ \hat x = \sum_{i} x_i |i\rangle\langle i|$  which satisfies \begin{align}
    \hat x \hat a &=\sum_{ij}  |i\rangle x_ia_{ij} \langle j|\\
    \hat a\hat x &=\sum_{ij}  |i\rangle a_{ij}x_j \langle j|\\
    \left[\hat x, \hat a\right]_{ij} &= (x_i - x_j) a_{ij}
\end{align} and define $\hat a^{\hat x} = -i\left[\hat x, \hat a\right] $. We can then write \begin{align}
    \sum_k \Phi_{ikj} t_{ik} \mathcal{G}_{kj} &= \dfrac{eB}{2\hbar}\sum_{k} \left[(x_i-x_k)(y_k-y_j) - (y_i - y_k)(x_k-x_j)\right] t_{ik} \mathcal{G}_{kj} \\
    &= \dfrac{eB}{2\hbar}\sum_{k}   (x_it_{ik}-t_{ik}x_k)(y_k\mathcal{G}_{kj}-\mathcal{G}_{kj}y_j)  - (y_it_{ik}  - t_{ik} y_k)(x_k\mathcal{G}_{kj}-\mathcal{G}_{kj}x_j)  \\
    &= \dfrac{eB}{2\hbar}\sum_{k} (\hat x\mathcal{H} - \mathcal{H}\hat x)_{ik} (\hat y\mathcal{G} - \mathcal{G}\hat y)_{kj} - (\hat y\mathcal{H}- \mathcal{H}\hat y)_{ik}(\hat x\mathcal{G} - \mathcal{G}\hat x)_{kj}\\
    &= \dfrac{eB}{2\hbar} \left[\left[\hat x,\mathcal{H}\right]\left[\hat y,\mathcal{G}\right] - \left[\hat y,\mathcal{H}\right]\left[\hat x,\mathcal{G}\right]\right]_{ij}\\
    \sum_k \Phi_{ikj} t_{ik} \mathcal{G}_{kj}&= - \dfrac{eB}{2\hbar} \left[\mathcal{H}^{\hat x} \mathcal{G}^{\hat y} - \mathcal{H}^{\hat y} \mathcal{G}^{\hat x}\right]_{ij}
\end{align} We get \begin{align}
    \mathcal{G}^{B}_{ii} &= \mathcal{G}_{ii}  +  i \sum_{j}   \mathcal{G}_{ij}  \sum_{k}\Phi_{jki} t_{jk}   \mathcal{G}_{ki} \\
    &=\mathcal{G}_{ii}  - \dfrac{ieB}{2\hbar}\sum_{j}  \mathcal{G}_{ij}\left[\mathcal{H}^{\hat x} \mathcal{G}^{\hat y} - \mathcal{H}^{\hat y} \mathcal{G}^{\hat x}\right]_{ji} \\
     \mathcal{G}^{B}_{ii} &= \mathcal{G}_{ii}  - \dfrac{ieB}{2\hbar} (\mathcal{G}\left[\mathcal{H}^{\hat x} \mathcal{G}^{\hat y} - \mathcal{H}^{\hat y} \mathcal{G}^{\hat x}\right])_{ii} 
    \label{eq:TrG1}
\end{align} 
We thus obtain \begin{align}
    \partial_B \mathcal{G}^B_{ii}|_{B = 0} = - \dfrac{ie}{2\hbar} (\mathcal{G}\left[\mathcal{H}^{\hat x} \mathcal{G}^{\hat y} - \mathcal{H}^{\hat y} \mathcal{G}^{\hat x}\right])_{ii} 
\end{align} 
Reintroducing the position variable $\bm r$ instead of index $i$ we obtain: \begin{align}
    M^z_{\text{orb}}(\bm r) &= \dfrac{e}{2\pi \hbar} \int_{-\infty}^{+\infty}\mathrm{d}\varepsilon F(\varepsilon) \mathrm{Re} \langle \bm r|  \mathcal{G}\left[\mathcal{H}^{\hat x} \mathcal{G}^{\hat y} - \mathcal{H}^{\hat y} \mathcal{G}^{\hat x}\right] | \bm r \rangle 
    \label{eq:MorbGreensZdirection}
\end{align} Considering all directions, the position operator is given by ${\hat {\bm r}}$ and the velocity operator is defined (up to $\hbar$) as ${\hat {\bm v}} = \mathcal{H}^{\hat {\bm r}} = -i \left[\hat {\bm r},\mathcal{H} \right]$. The magnetization in all directions is given by  \begin{align}
    \bm M_{\text{orb}}(\bm r) = \dfrac{e}{2\pi \hbar}  \int_{-\infty}^{+\infty}\mathrm{d}\varepsilon F(\varepsilon) \mathrm{Re}  \langle \bm r|  \mathcal{G} {\hat {\bm v}} \times \mathcal{G}^{\hat {\bm r}}  | \bm r \rangle 
    \label{eq:MorbGreenRspace}
\end{align} using the identity $\mathcal{G}^{\hat {\bm r}} = \mathcal{G}{\hat {\bm v}}\mathcal{G}$, it can also be expressed as 
\begin{align}
    \Aboxed{\bm M_{\text{orb}}(\bm r) = \dfrac{e}{2\pi \hbar}  \int_{-\infty}^{+\infty}\mathrm{d}\varepsilon F(\varepsilon) \mathrm{Re}  \langle \bm r|  \mathcal{G}(\hat {\bm v} \mathcal{G}) \times (\hat {\bm v} \mathcal{G}) | \bm r \rangle }
    \label{eq:MorbGreenArticle}
\end{align} which corresponds to equation (3) of the main article.

\section{Theory of local orbital magnetization}
\label{sec:ThermoTheory}

In this section, we compute the expression of the local orbital magnetization for the relevant geometries of the sample depending on the number of Periodic Boundary Conditions (PBC).
The simplest expression is obtained for finite size systems with full open boundary condition as described in the following $\bm r$-space section.
The case of Bloch electron is detailed in the $\bm k$-space section.
The last section describes the intermediate case of a ribbon geometry.

As a reminder, we define $f(\varepsilon) = 1/(1+e^{\beta(\varepsilon-\mu)})$ the fermi function and  $F(\varepsilon) = k_B T \ln\left(1 + e^{-\beta(\varepsilon - \mu)}\right)$ a thermodynamic weight which satisfies $F^\prime(\varepsilon) = - f(\varepsilon)$, with $\mu$ the chemical potential and $T$ the temperature. 

\subsection{r-space}
Finite-size systems are described by their hamiltonian $\mathcal{H}_{\bm r} = \mathcal{H}$ in position space ($\bm r$-space), and their associated eigenstates $|n\rangle$. Starting from equation \eqref{eq:MorbGreenRspace}, we have 
\begin{align}
    \bm M_{\text{orb}}(\bm r)  &= \dfrac{e}{2\pi \hbar} \int_{-\infty}^{+\infty}\mathrm{d}\varepsilon F(\varepsilon) \mathrm{Re} \langle \bm r|   \mathcal{G} {\hat {\bm v}} \times \mathcal{G}^{\hat {\bm r}} | \bm r \rangle 
\end{align}
We have \begin{align}
    \mathcal{G} {\hat {\bm v}} \times \mathcal{G}^{\hat {\bm r}} &= (-i)^2 \mathcal{G} ( {\hat {\bm r}}\mathcal{H} - \mathcal{H}{\hat {\bm r}} )\times ( {\hat {\bm r}}\mathcal{G} - \mathcal{G}{\hat {\bm r}} )\\
    &= -\mathcal{G}( {\hat {\bm r}}\mathcal{H} \times  {\hat {\bm r}}\mathcal{G}  - {\hat {\bm r}}\mathcal{H}  \times \mathcal{G}{\hat {\bm r}}  -  \mathcal{H} \underbrace{{\hat {\bm r}}  \times {\hat {\bm r}}}_{{\hat {\bm r}}  \times {\hat {\bm r}} = 0} \mathcal{G}  + \mathcal{H}{\hat {\bm r}} \times  \mathcal{G}{\hat {\bm r}} )\\
    &= -( \mathcal{G}{\hat {\bm r}}\mathcal{H} \times  {\hat {\bm r}}\mathcal{G}  - \mathcal{G} ({\hat {\bm r}}\mathcal{H} - \mathcal{H}{\hat {\bm r}})  \times \mathcal{G} {\hat {\bm r}}  )\\
    &= -( \mathcal{G}{\hat {\bm r}}\mathcal{H} \times  {\hat {\bm r}}\mathcal{G}  - \mathcal{G} ({\hat {\bm r}}\mathcal{H} - \mathcal{H}{\hat {\bm r}}) \mathcal{G} \times {\hat {\bm r}}  )\label{eq:greensreal}
\end{align} By definition of the Green's function, we have \begin{align}
    &\begin{cases}
        (\varepsilon - \mathcal{H})\mathcal{G} = 1\\
        \mathcal{G}(\varepsilon - \mathcal{H}) = 1
    \end{cases}\\
    &\begin{cases}
        \varepsilon\mathcal{G} - \mathcal{H}\mathcal{G} = 1\\
        \varepsilon\mathcal{G} - \mathcal{G}\mathcal{H} = 1
    \end{cases}\\
    &\begin{cases}
        \varepsilon\hat{\bm r}\mathcal{G} - \hat{\bm r}\mathcal{H}\mathcal{G} = \hat{\bm r}\\
        \varepsilon\mathcal{G}\hat{\bm r} - \mathcal{G}\mathcal{H}\hat{\bm r} = \hat{\bm r}
    \end{cases}\\
    &\begin{cases}
        \varepsilon\mathcal{G}\hat{\bm r}\mathcal{G} - \mathcal{G}\hat{\bm r}\mathcal{H}\mathcal{G} = \mathcal{G}\hat{\bm r}\\
        \varepsilon\mathcal{G}\hat{\bm r}\mathcal{G} - \mathcal{G}\mathcal{H}\hat{\bm r}\mathcal{G} = \hat{\bm r}\mathcal{G}
    \end{cases}
\end{align} This gives us \begin{align}
    \mathcal{G} \left[\hat{\bm r}, \mathcal{H}\right]\mathcal{G} = \left[\hat{\bm r}, \mathcal{G}\right]
    \label{eq:commutationHG}
\end{align} and hence, going back to equation \eqref{eq:greensreal}, we have \begin{align}
    \mathcal{G} {\hat {\bm v}} \times \mathcal{G}^{\hat {\bm r}}= - \mathcal{G}({\hat {\bm r}}\mathcal{H} \times {\hat {\bm r}}  )\mathcal{G}  + {\hat {\bm r}}\mathcal{G} \times{\hat {\bm r}} 
\end{align}
We thus have \begin{align}
    \langle \bm r|  \mathcal{G} {\hat {\bm v}} \times \mathcal{G}^{\hat {\bm r}} | \bm r \rangle  &=  - \langle \bm r|\mathcal{G}({\hat {\bm r}}\mathcal{H} \times {\hat {\bm r}}  )\mathcal{G}| \bm r \rangle + \langle \bm r|{\hat {\bm r}}\mathcal{G} \times{\hat {\bm r}} | \bm r \rangle \\
    &=-\sum_{nm} \dfrac{\langle \bm r|n\rangle \langle m|\bm r \rangle ({\hat {\bm r}}\mathcal{H} \times {\hat {\bm r}})_{nm}}{(\varepsilon - \varepsilon_n)(\varepsilon - \varepsilon_m)} + \sum_{n} \dfrac{|\langle \bm r|n\rangle|^2}{\varepsilon - \varepsilon_n}  \underbrace{{\bm r} \times {\bm r}}_{ = 0}\\
    \langle \bm r|  \mathcal{G} {\hat {\bm v}} \times \mathcal{G}^{\hat {\bm r}} | \bm r \rangle&=-\sum_{nm} \dfrac{\langle \bm r|n\rangle \langle m|\bm r \rangle ({\hat {\bm r}}\mathcal{H} \times {\hat {\bm r}})_{nm}}{(\varepsilon - \varepsilon_n)(\varepsilon - \varepsilon_m)} 
\end{align}
This gives us \begin{align}
    \bm M_{\text{orb}}(\bm r) &=-\dfrac{e}{2\pi \hbar} \int_{-\infty}^{+\infty}\mathrm{d}\varepsilon F(\varepsilon) \mathrm{Re} \sum_{nm} \dfrac{\langle \bm r|n\rangle \langle m|\bm r \rangle ({\hat {\bm r}}\mathcal{H} \times {\hat {\bm r}})_{nm}}{(\varepsilon - \varepsilon_n)(\varepsilon - \varepsilon_m)} \\
    &= \dfrac{e}{2\pi \hbar}\sum_{nm}\mathrm{Im} \left\{ {\langle \bm r|n\rangle \langle m|\bm r \rangle ({\hat {\bm r}}\mathcal{H} \times {\hat {\bm r}})_{nm}} \right\} \int_{-\infty}^{+\infty}\mathrm{d}\varepsilon\mathrm{Im} \dfrac{ F(\varepsilon) }{(\varepsilon - \varepsilon_n)(\varepsilon - \varepsilon_m)} 
\end{align}
But \begin{align}
    \mathrm{Im} \dfrac{1 }{(\varepsilon - \varepsilon_n)^2}  &= -\pi  \delta^\prime(\varepsilon - \varepsilon_n) \\
    \mathrm{Im} \dfrac{1 }{(\varepsilon - \varepsilon_n)(\varepsilon - \varepsilon_m)}  &= - \pi \dfrac{\delta(\varepsilon - \varepsilon_n)-\delta(\varepsilon - \varepsilon_m)}{\varepsilon_n - \varepsilon_m}
\end{align} We end up with \begin{align}
    \bm M_{\text{orb}}(\bm r)&= \dfrac{e}{2\hbar} \sum_n f(\varepsilon_n) |\langle \bm r|n\rangle|^2 \mathrm{Im}  \left\{ ({\hat {\bm r}}\mathcal{H} \times {\hat {\bm r}})_{nn} \right\} - \dfrac{e}{\hbar}\sum_{n}\sum_{m \ne n} \dfrac{F(\varepsilon_n)}{\varepsilon_n - \varepsilon_m} \mathrm{Im} \left\{ \langle \bm r|n\rangle \langle m|\bm r \rangle ({\hat {\bm r}}\mathcal{H} \times {\hat {\bm r}})_{nm} \right\}
\end{align}
We recognize $ {\hat {\bm r}} \mathcal{H} \times {\hat {\bm r}}= {\hat {\bm r}}  \times \mathcal{H} {\hat {\bm r}} = - {\hat {\bm r}} \times ({\hat {\bm r}}\mathcal{H} - \mathcal{H}{\hat {\bm r}}) = - i {\hat {\bm r}} \times  {\hat {\bm v}}  $. This gives us \begin{align}
    \bm M_{\text{orb}}(\bm r)&=  \dfrac{e}{\hbar} \sum_n  \left\{- \dfrac{1}{2}f(\varepsilon_n)|\langle \bm r|n\rangle|^2    ({\hat {\bm r}} \times  {\hat {\bm v}})_{nn} +  \sum_{m \ne n} \dfrac{F(\varepsilon_n)}{\varepsilon_n - \varepsilon_m} \mathrm{Re} \left\{ \langle \bm r|n\rangle \langle m|\bm r \rangle ({\hat {\bm r}} \times {\hat {\bm v}})_{nm} \right\}\right\}
\end{align}
We obtain the general definition discussed in the main article \begin{align}
    \Aboxed{{\bm M}_{\textrm{orb}}({\bm r}) = \dfrac{e}{\hbar} \sum_n \left[f(\varepsilon_n)  {\bm m}_{n}(\bm r)  + F(\varepsilon_n) {\bm \Omega}_n(\bm r) \right]}
    \label{eq:MorbFinsize}
\end{align}   with 
\begin{align}  
{\bm m}_{n}(\bm r)= - \dfrac{1}{2}|\langle \bm r|n\rangle|^2    ({\hat {\bm r}} \times  {\hat {\bm v}})_{nn} =  |\langle \bm r|n\rangle|^2 {\bm m}_{n}
\label{eq:LocalBerry}\end{align}   the local orbital momentum term with ${\bm m}_n= -\frac{1}{2} \langle n| {\hat {\bm r}}\times  {\hat {\bm v}}|n\rangle$  the orbital momentum of state $|n\rangle$   and 
\begin{align}  
\bm \Omega_n(\bm r) &= \mathrm{Re}  \sum_{m \ne n} \langle \bm r|n\rangle \langle m|\bm r \rangle\dfrac{   \langle n| {\hat {\bm r}}\times  {\hat {\bm v}}|m\rangle}{\varepsilon_n - \varepsilon_m}   \label{eq:LocalBerry1}
\end{align} the local Berry curvature term. The local Chern number in $\bm r$-space can be defined accordingly as \begin{align}
    \bm C(\bm r) = 2\pi \sum_{n} f(\varepsilon_n) \bm \Omega_n(\bm r)
\end{align}

\subsection{k-space}

We consider now the case of extended systems described by a Bloch hamiltonian $\mathcal{H}_{\bm k}$. We have ${\bm v}_{\bm k}  = \partial_{\bm k} \mathcal{H}_{\bm k}$. We start with equation \eqref{eq:MorbGreenArticle}. The local orbital magnetization is defined as \begin{align}
    \bm M_{\text{orb}}(\bm r) = \dfrac{e}{2\pi \hbar}  \int_{-\infty}^{+\infty}\mathrm{d}\varepsilon F(\varepsilon) \mathrm{Re}  \langle \bm r|  \mathcal{G}(\hat {\bm v} \mathcal{G}) \times ( \hat{\bm v} \mathcal{G}) | \bm r \rangle
\end{align} where now $\bm r$  corresponds to the position of one of the sublattices within the unit cell. Using the notation $\langle \bullet \rangle_{\bm k} =\int \frac{d^2k}{(2\pi)^2} \bullet$ from the main article, we have \begin{align}
    &\bm M_{\text{orb}}(\bm r) = \dfrac{e}{2\pi \hbar} \int_{-\infty}^{+\infty}\mathrm{d}\varepsilon F(\varepsilon) \mathrm{Re} \sum_{nml} \left\langle \dfrac{ \langle \bm r| n\bm k\rangle  \langle l \bm k| \bm r \rangle {\bm v}_{nm\bm k} \times   {\bm v}_{ml\bm k} }{(\varepsilon - \varepsilon_{n\bm k})(\varepsilon - \varepsilon_{m\bm k})(\varepsilon - \varepsilon_{l\bm k})} \right\rangle _{\bm k} 
    \\
    &= - \dfrac{e}{2\pi \hbar} \sum_{nml} \left\langle  \mathrm{Im}\left\{ \langle \bm r| n\bm k\rangle  \langle l \bm k| \bm r \rangle {\bm v}_{nm\bm k} \times   {\bm v}_{ml\bm k} \right\} \mathrm{Im} \int_{-\infty}^{+\infty}  \dfrac{F(\varepsilon)\mathrm{d}\varepsilon}{(\varepsilon - \varepsilon_{n\bm k})(\varepsilon - \varepsilon_{m\bm k})(\varepsilon - \varepsilon_{l\bm k})}\right\rangle _{\bm k} 
\end{align} where we have used the notation $ {\bm v}_{nm\bm k} = \langle n\bm k | \partial_{\bm k} \mathcal{H}_{\bm k} | m\bm k\rangle $ for the interband velocity. 
As we did in the previous part, we would like to perform the energy integral. We first perform the partial fraction decomposition given by  \begin{align}
    \dfrac{1}{(\varepsilon - \varepsilon_{n})(\varepsilon - \varepsilon_{m})(\varepsilon - \varepsilon_{l})} &= \dfrac{1}{(\varepsilon_{n} - \varepsilon_{m})(\varepsilon_{n} - \varepsilon_{l})(\varepsilon_{m} - \varepsilon_{l})} \left( \dfrac{\varepsilon_{m} - \varepsilon_{l}}{\varepsilon - \varepsilon_{n}} + \dfrac{\varepsilon_{l} - \varepsilon_{n}}{\varepsilon - \varepsilon_{m}}  + \dfrac{\varepsilon_{n} - \varepsilon_{m}}{\varepsilon - \varepsilon_{l}}   \right) \\
    \dfrac{1}{(\varepsilon - \varepsilon_{n})^2(\varepsilon - \varepsilon_{m})} &= \dfrac{1}{\varepsilon_{n}-\varepsilon_{m}}\dfrac{1}{(\varepsilon-\varepsilon_{n})^2}  + \dfrac{1}{(\varepsilon_{n}-\varepsilon_{m})^2} \left(\dfrac{1}{\varepsilon-\varepsilon_{m}} - \dfrac{1}{\varepsilon-\varepsilon_{n}}\right) 
\end{align} and perform the integral with the knowledge of the following identities \begin{align}
    \mathrm{Im} \dfrac{1}{(\varepsilon - \alpha + i\eta)^n} &= - \dfrac{\pi}{(n-1)!} \delta^{(n-1)} (\varepsilon - \alpha)
\end{align} We thus have \begin{align}
    \mathrm{Im} \dfrac{1}{(\varepsilon - \varepsilon_{n})^2(\varepsilon - \varepsilon_{m})} &= - \dfrac{ \pi \delta^\prime(\varepsilon - \varepsilon_{n})}{\varepsilon_{n}-\varepsilon_{m}} - \dfrac{\pi }{(\varepsilon_{n}-\varepsilon_{m})^2} \left(\delta(\varepsilon-\varepsilon_{m}) - \delta(\varepsilon-\varepsilon_{n})\right) 
\end{align}
 For simplicity we define \begin{align}
    \mathcal{M}^{\bm r }_{nml} &=  \langle \bm r| n\bm k\rangle  \langle l \bm k| \bm r \rangle {\bm v}_{nm\bm k} \times   {\bm v}_{ml\bm k} 
\end{align}
We thus have \begin{align}
   \bm M_{\text{orb}}(\bm r) = - \dfrac{e}{2\pi \hbar}    \sum_{nml}  \left\langle\mathrm{Im}(\mathcal{M}^{\bm r }_{nml} )\int_{-\infty}^{+\infty}\mathrm{d}\varepsilon F(\varepsilon) \mathrm{Im}\dfrac{1}{(\varepsilon - \varepsilon_{n\bm k})(\varepsilon - \varepsilon_{m\bm k})(\varepsilon - \varepsilon_{l\bm k})}\right\rangle_{\bm k}
\end{align} We have $\mathcal{M}^{\bm r }_{nnn} = 0$ identically. We can expand the sum, and we get \begin{align}
    \bm M_{\text{orb}}(\bm r) = &-  \dfrac{e}{2\pi \hbar}   \sum_{n}\sum_{m \ne n}  \left\langle \mathrm{Im}(\mathcal{M}^{\bm r}_{nmn} + \mathcal{M}^{\bm r}_{nnm}+ \mathcal{M}^{\bm r}_{mnn}) \mathrm{Im}\int_{-\infty}^{+\infty} \dfrac{F(\varepsilon)\mathrm{d}\varepsilon }{(\varepsilon - \varepsilon_{n\bm k})^2(\varepsilon - \varepsilon_{m\bm k})} \right\rangle _{\bm k} \nonumber \\
    &-  \dfrac{e}{2\pi \hbar} \sum_{n}\sum_{m \ne n}\sum_{l \ne m,n}\left\langle  \mathrm{Im} (\mathcal{M}^{\bm r}_{nml})\mathrm{Im} \int_{-\infty}^{+\infty} \dfrac{F(\varepsilon)\mathrm{d}\varepsilon }{(\varepsilon - \varepsilon_{n\bm k})(\varepsilon - \varepsilon_{m\bm k})(\varepsilon - \varepsilon_{l\bm k})}  \right\rangle_{\bm k}
\end{align}  We notice that $(\mathcal{M}^{\bm r }_{nml})^* = - \mathcal{M}^{\bm r }_{lmn}$ so that $\mathrm{Im} \mathcal{M}^{\bm r}_{lmn} = \mathrm{Im} \mathcal{M}^{\bm r}_{nml}$. We thus get \begin{align}
    &\bm M_{\text{orb}}(\bm r)=  -   \dfrac{e}{2 \hbar}\sum_{n}\sum_{m \ne n} \left\langle \mathrm{Im} (\mathcal{M}^{\bm r}_{nmn} + 2\mathcal{M}^{\bm r}_{nnm})  \left( \dfrac{f(\varepsilon_{n\bm k})}{\varepsilon_{n\bm k}-\varepsilon_{m\bm k}} + \dfrac{F(\varepsilon_{n\bm k}) - F(\varepsilon_{m\bm k})}{(\varepsilon_{n\bm k}-\varepsilon_{m\bm k})^2}\right)\right\rangle _{\bm k}   \\
    & +   \dfrac{e}{2 \hbar}  \sum_{n}\sum_{m \ne n}\sum_{l \ne m,n} \left\langle \mathrm{Im}( \mathcal{M}^{\bm r}_{nml} )
    \dfrac{(\varepsilon_{m\bm k} - \varepsilon_{l\bm k})F(\varepsilon_{n\bm k}) + (\varepsilon_{l\bm k} - \varepsilon_{n\bm k})F(\varepsilon_{m\bm k})  + (\varepsilon_{n\bm k} - \varepsilon_{m\bm k})F(\varepsilon_{l\bm k})}{(\varepsilon_{n\bm k} - \varepsilon_{m\bm k})(\varepsilon_{n\bm k} - \varepsilon_{l\bm k})(\varepsilon_{m\bm k} - \varepsilon_{l\bm k})}  \right\rangle_{\bm k}\nonumber
\end{align}
We can proceed further by recasting all the Fermi functions as functions of $\varepsilon_{n}$. We get \begin{align}
    \bm M_{\text{orb}}(\bm r)= 
    &-   \dfrac{e}{2 \hbar}   \mathrm{Im}\sum_{n}\sum_{m \ne n}   \left\langle \dfrac{  \mathcal{M}^{\bm r}_{nmn} + 2\mathcal{M}^{\bm r}_{nnm}}{\varepsilon_{n\bm k}-\varepsilon_{m\bm k}} f(\varepsilon_{n\bm k}) + \dfrac{\mathcal{M}^{\bm r}_{nmn}  +2 \mathcal{M}^{\bm r}_{nnm} - \mathcal{M}^{\bm r}_{mnm} -  2\mathcal{M}^{\bm r}_{nmm}}{(\varepsilon_{n\bm k}-\varepsilon_{m\bm k})^2}F(\varepsilon_{n\bm k}) \right\rangle _{\bm k} \nonumber\\
    & +   \dfrac{e}{2 \hbar} \mathrm{Im}\sum_{n}\sum_{m \ne n}\sum_{l \ne m,n}  \left\langle  
    \dfrac{ \mathcal{M}^{\bm r}_{mnl}  + 2\mathcal{M}^{\bm r}_{nml} }{(\varepsilon_{n\bm k} - \varepsilon_{m\bm k})(\varepsilon_{n\bm k} - \varepsilon_{l\bm k})} F(\varepsilon_{n\bm k})\right\rangle _{\bm k}
\end{align} We define ${\bm v}_{n\bm k} = {\bm v}_{n n\bm k}$ the intraband velocity and  ${\bm r}_{n m\bm k}={\bm v}_{n m\bm k}/(\varepsilon_{n\bm k}-\varepsilon_{m\bm k})$ the interband position. We obtain
\begin{align}
    \bm M_{\text{orb}}&(\bm r)= 
    \dfrac{e}{2\hbar}   \mathrm{Im}\sum_{n}\sum_{m \ne n}   \left\langle \left\{ -|\langle \bm r| n\bm k\rangle|^2 {\bm r}_{nm\bm k} \times   {\bm v}_{mn\bm k}  + 2\langle \bm r| n\bm k\rangle  \langle m \bm k| \bm r \rangle {\bm r}_{nm\bm k}\times {\bm v}_{n\bm k} \right\} f(\varepsilon_{n\bm k})\right\rangle _{\bm k} \\
    &+\dfrac{e}{2 \hbar}   \mathrm{Im}\sum_{n}\sum_{m \ne n}   \left\langle  \left\{|\langle \bm r| n\bm k\rangle|^2 {\bm r}_{nm\bm k} \times   {\bm r}_{mn\bm k}  +2 \dfrac{ \langle \bm r| n\bm k\rangle  \langle m \bm k| \bm r \rangle}{\varepsilon_{n\bm k}-\varepsilon_{m\bm k}} {\bm r}_{nm\bm k}\times {\bm v}_{n\bm k} \right\}F(\varepsilon_{n\bm k}) \right\rangle _{\bm k} \nonumber\\
    &+  \dfrac{e}{2 \hbar}   \mathrm{Im}\sum_{n}\sum_{m \ne n}   \left\langle \left\{|\langle \bm r| m\bm k\rangle|^2 {\bm r}_{nm\bm k} \times   {\bm r}_{mn\bm k}  + 2\dfrac{\langle \bm r| m\bm k\rangle  \langle n \bm k| \bm r \rangle}{\varepsilon_{n\bm k}-\varepsilon_{m\bm k}}{\bm r}_{mn\bm k} \times {\bm v}_{m\bm k}\right\}F(\varepsilon_{n\bm k}) \right\rangle _{\bm k} \nonumber\\
    & +   \dfrac{e}{2 \hbar} \mathrm{Im}\sum_{n}\sum_{m \ne n}\sum_{l \ne m,n}  \left\langle   \left\{
    - \langle \bm r| m\bm k\rangle  \langle l \bm k| \bm r \rangle {\bm r}_{mn\bm k} \times   {\bm r}_{nl\bm k}  + 2\dfrac{\langle \bm r| n\bm k\rangle  \langle l \bm k| \bm r \rangle }{\varepsilon_{n\bm k} - \varepsilon_{l\bm k}}{\bm r}_{nm\bm k} \times   {\bm v}_{ml\bm k} \right\}F(\varepsilon_{n\bm k})\right\rangle _{\bm k}\nonumber
\end{align}
Using integration by parts on the second term of the first line, we have 
{\small \begin{align}
    &\left\langle f(\varepsilon_{n\bm k}) \sum_{m \ne n}   \langle \bm r| n\bm k\rangle  \langle m \bm k| \bm r \rangle {\bm r}_{nm\bm k}\times {\bm v}_{n\bm k} \right\rangle _{\bm k} =\left\langle F(\varepsilon_{n\bm k}) \sum_{m \ne n} \left(|\langle \bm r |n \rangle |^2 -|\langle \bm r | m \rangle|^2 \right) {\bm r}_{nm \bm k} \times {\bm r}_{mn \bm k}  \right.\\
    & +\left.  F(\varepsilon_{n\bm k}) \sum_{m \ne n} \sum_{l \ne n,m}  \left[ \left( \langle \bm r | l \rangle \langle m  | \bm r \rangle {\bm r}_{ln \bm k} \times {\bm r}_{nm \bm k} +\langle \bm r |n \rangle  \langle  l | \bm r \rangle {\bm r}_{nm \bm k} \times {\bm r}_{ml \bm k}\right)- \dfrac{\langle \bm r |n \rangle \langle m| \bm r \rangle}{\varepsilon_{n\bm k}-\varepsilon_{m\bm k} }\left({\bm r}_{nl \bm k}\times{\bm v}_{lm \bm k} + {\bm v}_{nl \bm k}\times{\bm r}_{lm \bm k} \right)\right]\right\rangle _{\bm k} 
\end{align}}
Collecting all the terms, we thus obtain \begin{align}
    \Aboxed{
    {\bm M}_{\textrm{orb}}(\bm r) &=  \dfrac{e}{\hbar}  \sum_{n} \left\langle  f(\varepsilon_{n\bm k}) {\bm m}_{n \bm k}(\bm r) + F(\varepsilon_{n \bm k}) {\bm \Omega}_{n \bm k}(\bm r)\right\rangle _{\bm k} }
    \label{eq:Morbkspace}
\end{align} with \begin{align}
    {\bm m}_{n \bm k}(\bm r) = &|\langle \bm r| n \bm k \rangle|^2{\bm m}_{n \bm k}\\
    {\bm \Omega}_{n \bm k}(\bm r) = & |\langle \bm r| n \bm k \rangle|^2  {\bm \Omega}_{n \bm k} + {\bm \Omega}^{\text{geom}}_{n \bm k}(\bm r)
\end{align} and ${\bm \Omega}^{\text{geom}}_{n \bm k}(\bm r)$ given for many-band systems by \begin{align}
    &{\bm \Omega}^{\text{geom}}_{n \bm k}(\bm r)= \dfrac{1}{2} |\langle \bm r |n \rangle |^2  {\bm \Omega}_{n\bm k} - \dfrac{1}{2}  \sum_{m \ne n} |\langle \bm r | m \rangle|^2 {\bm \Omega}_{n m\bm k}+ \sum_{m \ne n} \mathrm{Im} \dfrac{\langle \bm r| n \bm k \rangle \langle m \bm k| \bm r \rangle}{\varepsilon_{n \bm k}-\varepsilon_{m \bm k}} {\bm r}_{nm \bm k}\times ({\bm v}_{n \bm k} + {\bm v}_{m \bm k}) \\
    &+ \mathrm{Im} \sum_{m \ne n} \dfrac{\langle \bm r |n \rangle \langle m| \bm r \rangle}{\varepsilon_{n\bm k}-\varepsilon_{m\bm k} } \sum_{l \ne n,m}   {\bm r}_{lm \bm k} \times {\bm v}_{nl \bm k}+ \mathrm{Im} \sum_{m \ne n} \sum_{l \ne n,m}  \left[ \dfrac{1}{2} \langle \bm r | l \rangle \langle m  | \bm r \rangle {\bm r}_{ln \bm k} \times {\bm r}_{nm \bm k} +\langle \bm r |n \rangle  \langle  l | \bm r \rangle {\bm r}_{nm \bm k} \times {\bm r}_{ml \bm k} \right]\nonumber
\end{align} where ${\bm m}_{n \bm k}= -\frac{1}{2}{\bm r}_{nm \bm k} \times {\bm v}_{mn \bm k}$ is the orbital momentum, ${\bm \Omega}_{n m\bm k} = {\bm r}_{nm \bm k} \times {\bm r}_{mn \bm k}$ the interband Berry curvature and ${\bm \Omega}_{n \bm k} = \sum_{m\ne n}{\bm \Omega}_{n m\bm k}$ the usual intraband Berry curvature.

\subsection*{Application: two-band models}

We consider in this section a extended two-band model corresponding to a lattice with two non equivalent sites per unit cell. In reciprocal space, this system is generically described by a Hamiltonian of the form \begin{align}
    \mathcal{H}({\bm k}) = \varepsilon_0(\bm k)\textrm{I} + \bm d({\bm k}) \cdot \bm \sigma 
\end{align} We denote $n = \pm$ the two bands, such that $\varepsilon_{n, \bm k} = \varepsilon_0 + n |\bm d|$.  
We also denote $P_{\bm r}=\frac{1}{2}(1+\bm s_{\bm r}\cdot \bm \sigma )$
the sublattice projector, with $\bm s_{\bm r}= + {\bm e}_z$ for $A$ sublattice and $s_{\bm r}=-{\bm e}_z$ for $B$ sublattice. We introduce the unit vector $\hat {\bm d} = \bm d/|\bm d|$ and the antisymetric tensor $\eta_{\alpha \beta \gamma}$ where  $\alpha, \beta, \gamma \in \{x, y, z\}$. We then obtain the $z$-component of the sublattice orbital momentum and sublattice Berry curvature  as \begin{align}
    m_{n ,\bm k}^\gamma(\bm r)&=
    -\frac{\eta_{\alpha \beta \gamma}}{8}(1+n \bm s_{\bm r} \cdot \hat {\bm d})\bm d  \cdot \left(\partial_{k_\alpha} \hat {\bm d} \times \partial_{k_\beta} \hat {\bm d}\right) \\
    {\Omega}^{\gamma}_{n \bm k}(\bm r)&= - \eta_{\alpha \beta \gamma}\dfrac{n}{8}\hat {\bm d} \cdot \left(\partial_{k_\alpha} \hat {\bm d} \times \partial_{k_\beta} \hat {\bm d}\right) + {\Omega}^{\text{geom},\gamma}_{n \bm k}(\bm r)\\
    {\Omega}^{\text{geom},\gamma}_{n \bm k}(\bm r)&= - \eta_{\alpha \beta \gamma} \dfrac{n}{4}\bm s_{\bm r}\cdot \left( \left[\dfrac{\partial_{k_\alpha} \varepsilon_0 }{|\bm d|}\hat {\bm d}  + n \partial_{k_\alpha} \hat {\bm d}  \right] \times \partial_{k_\beta} \hat {\bm d} \right)
\end{align} 
As discussed in the main article, and made explicit by the last expression, in two-band models the non-quantized contribution of ${\Omega}^{\text{geom},\gamma}_{n \bm k}(\bm r)$ to the slope of the sublattice orbital magnetization in the gap is non-zero only for non particle-hole symmetric system ($\varepsilon_0 \ne 0$), as the second term in ${\Omega}^{\text{geom},\gamma}_{n \bm k}(\bm r)$ integrates to zero over an occupied band. 

\subsection{ribbon geometry}
\label{sec:ribbon}

We consider the case of a sample with a cylindrical geometry, meaning that the system is infinite is the $x$ direction and finite in the second direction $y$ with open boundary conditions in the finite direction. We use the notation $|n,k_x\rangle = |n\rangle$ omitting the index $k_x$ in the notation.

Once again, we start with the Green's function formulation of the local orbital magnetization given by equation \eqref{eq:MorbGreensZdirection}. We have \begin{align}
    M^z_{\text{orb}}(\bm r) &= \dfrac{e}{2\pi \hbar} \int_{-\infty}^{+\infty}\mathrm{d}\varepsilon F(\varepsilon) \mathrm{Re} \langle \bm r|  \mathcal{G}\left[\mathcal{H}^{\hat x} \mathcal{G}^{\hat y} - \mathcal{H}^{\hat y} \mathcal{G}^{\hat x}\right] | \bm r \rangle 
\end{align} In this context, the operator $\hat y$ is a well defined finite-size position operator and we have $\mathcal{H}^{\hat y} = -i [\hat y, \mathcal{H}]$. In the periodic $x$-direction, we have $\mathcal{H}^{\hat x} = \partial_{k_x} \mathcal{H} = v_{k_x}$ and consequently $\mathcal{G}^{\hat x}= \mathcal{G}v_{k_x} \mathcal{G}$. We thus have 
\begin{align}
     \mathcal{G}\left[\mathcal{H}^{\hat x} \mathcal{G}^{\hat y} - \mathcal{H}^{\hat y} \mathcal{G}^{\hat x}\right]  &= -i \left[\mathcal{G}v_{k_x}(\hat y\mathcal{G}-\mathcal{G}\hat y) - \mathcal{G}(\hat y\mathcal{H}-\mathcal{H} \hat y)\mathcal{G}^{\hat x}\right]\\
    &= -i \left[\mathcal{G}v_{k_x}(\hat y\mathcal{G}-\mathcal{G}\hat y) - \mathcal{G}(\hat y\mathcal{H}-\mathcal{H} \hat y)\mathcal{G}v_{k_x} \mathcal{G}\right]\\
\end{align} This gives us 
\begin{align}
    \mathrm{Re}  \langle \bm r|   \mathcal{G}&\left[\mathcal{H}^{\hat x} \mathcal{G}^{\hat y} - \mathcal{H}^{\hat y} \mathcal{G}^{\hat x}\right] | \bm r \rangle = \mathrm{Im}  \langle \bm r|  \left[\mathcal{G}v_{k_x}(\hat y\mathcal{G}-\mathcal{G}\hat y) - \mathcal{G}(\hat y\mathcal{H}-\mathcal{H} \hat y)\mathcal{G}v_{k_x} \mathcal{G}\right] | \bm r \rangle \\
    &= \mathrm{Im} \sum_{n,m} \dfrac{\langle \bm r|n\rangle \langle m|\bm r\rangle (v_{k_x}(\hat y-y_{\bm r}))_{nm}}{(\varepsilon - \varepsilon_n)(\varepsilon - \varepsilon_m)} 
    + \mathrm{Im} \sum_{n,m,l} \dfrac{\langle \bm r|n\rangle \langle l|\bm r\rangle (\varepsilon_n - \varepsilon_m)y_{nm}v_{ml ,k_x}}{(\varepsilon - \varepsilon_n)(\varepsilon - \varepsilon_m)(\varepsilon - \varepsilon_l)} \\
    &= \sum_{n,m}\mathrm{Re} \big\{\langle \bm r|n\rangle \langle m|\bm r\rangle (v_{k_x}(\hat y-y_{\bm r}))_{nm}\big\} \mathrm{Im} \dfrac{1}{(\varepsilon - \varepsilon_n)(\varepsilon - \varepsilon_m)}  \nonumber \\
    & \phantom{some blank}+ \sum_{n,m,l}\mathrm{Re}\big\{\langle \bm r|n\rangle \langle l|\bm r\rangle (\varepsilon_n - \varepsilon_m)y_{nm}v_{ml ,k_x}\big\} \mathrm{Im} \dfrac{1}{(\varepsilon - \varepsilon_n)(\varepsilon - \varepsilon_m)(\varepsilon - \varepsilon_l)} 
\end{align}  where $y_{\bm r} = \langle \bm r | \hat y | \bm r \rangle$ the position of state $|\bm r\rangle$ in the longitudinal unit-cell, $y_{nm} = \langle  n | \hat y |  m \rangle$ the position operator in the eigen basis, and $v_{nm ,k_x} = \langle n| v_{k_x}|m\rangle$ the velocity operator. From the previous computations in $\bm r$-space and $\bm k$-space, we have \begin{align}
    \mathrm{Im} \int_{-\infty}^{+\infty}\mathrm{d}\varepsilon \dfrac{F(\varepsilon)}{(\varepsilon - \varepsilon_n)^2}  &= \pi f(\varepsilon_n) \\
    \mathrm{Im} \int_{-\infty}^{+\infty}\mathrm{d}\varepsilon \dfrac{F(\varepsilon)}{(\varepsilon - \varepsilon_n)(\varepsilon - \varepsilon_m)}  &= -\pi \dfrac{F(\varepsilon_n) - F(\varepsilon_m)}{\varepsilon_n - \varepsilon_m}\\
    \mathrm{Im} \int_{-\infty}^{+\infty}\mathrm{d}\varepsilon \dfrac{F(\varepsilon)}{(\varepsilon - \varepsilon_n)^2(\varepsilon - \varepsilon_m)} &= \pi  \dfrac{ f(\varepsilon_n)}{\varepsilon_{n}-\varepsilon_{m}} + \pi \dfrac{ F(\varepsilon_n) - F(\varepsilon_m)}{(\varepsilon_{n}-\varepsilon_{m})^2}   \\
    \mathrm{Im} \int_{-\infty}^{+\infty}\mathrm{d}\varepsilon \dfrac{F(\varepsilon)}{(\varepsilon - \varepsilon_n)(\varepsilon - \varepsilon_m)(\varepsilon - \varepsilon_l)} &= -\pi \dfrac{(\varepsilon_{m} - \varepsilon_{l}) F(\varepsilon_n) + (\varepsilon_{l} - \varepsilon_{n})F(\varepsilon_m) + (\varepsilon_{n} - \varepsilon_{m}) F(\varepsilon_l) }{(\varepsilon_{n} - \varepsilon_{m})(\varepsilon_{n} - \varepsilon_{l})(\varepsilon_{m} - \varepsilon_{l})} 
\end{align}
We obtain the local orbital magnetization in the ribbon geometry \begin{align}
    &M^z_{\text{orb}}(\bm r) = \dfrac{e}{2 \hbar} \int_{-\pi}^{\pi} \dfrac{\mathrm{d}k_x}{2\pi} \sum_n  f(\varepsilon_n) \mathrm{Re} \big\{\langle \bm r|n\rangle \langle n|\bm r\rangle (v_{k_x}(\hat y-y_{\bm r}))_{nn}\big\} \nonumber\\
    &-\dfrac{e}{2 \hbar} \int_{-\pi}^{\pi} \dfrac{\mathrm{d}k_x}{2\pi} \sum_{n \ne m} \dfrac{F(\varepsilon_n) - F(\varepsilon_m)}{\varepsilon_n - \varepsilon_m} \mathrm{Re} \big\{\langle \bm r|n\rangle \langle m|\bm r\rangle (v_{k_x}(\hat y-y_{\bm r}))_{nm}\big\}\nonumber  \\
    &+ \dfrac{e}{2 \hbar}\int_{-\pi}^{\pi} \dfrac{\mathrm{d}k_x}{2\pi} \sum_{n \ne m} ( \dfrac{ f(\varepsilon_n)}{\varepsilon_{n}-\varepsilon_{m}} +  \dfrac{ F(\varepsilon_n) - F(\varepsilon_m)}{(\varepsilon_{n}-\varepsilon_{m})^2}) \mathrm{Re}\big\{\langle \bm r|n\rangle \langle n|\bm r\rangle (\varepsilon_n - \varepsilon_m)y_{nm}v_{mn,k_x}\big\}\nonumber \\
    &+ \dfrac{e}{2 \hbar}\int_{-\pi}^{\pi} \dfrac{\mathrm{d}k_x}{2\pi} \sum_{n \ne m} ( \dfrac{ f(\varepsilon_n)}{\varepsilon_{n}-\varepsilon_{m}} +  \dfrac{ F(\varepsilon_n) - F(\varepsilon_m)}{(\varepsilon_{n}-\varepsilon_{m})^2}) \mathrm{Re}\big\{\langle \bm r|m\rangle \langle n|\bm r\rangle (\varepsilon_m - \varepsilon_n)y_{mn}v_{nn,k_x}\big\} \\
    &-\dfrac{e}{2 \hbar}\int_{-\pi}^{\pi} \dfrac{\mathrm{d}k_x}{2\pi} \sum_{n \ne m \ne l}\dfrac{(\varepsilon_{m} - \varepsilon_{l}) F(\varepsilon_n) + (\varepsilon_{l} - \varepsilon_{n})F(\varepsilon_m) + (\varepsilon_{n} - \varepsilon_{m}) F(\varepsilon_l) }{(\varepsilon_{n} - \varepsilon_{m})(\varepsilon_{n} - \varepsilon_{l})(\varepsilon_{m} - \varepsilon_{l})} \mathrm{Re}\big\{\langle \bm r|n\rangle \langle l|\bm r\rangle (\varepsilon_n - \varepsilon_m)y_{nm}v_{ml,k_x}\big\}  \nonumber
\end{align}  It can be simplified into \begin{align}
    M^z_{\text{orb}}&(\bm r) = \dfrac{e}{2 \hbar}\int_{-\pi}^{\pi} \dfrac{\mathrm{d}k_x}{2\pi}  \sum_n  f(\varepsilon_n) \mathrm{Re} \big\{ |\langle \bm r|n\rangle|^2 (v_{k_x}(\hat y-y_{\bm r}))_{nn}\big\} \nonumber\\
    &-\dfrac{e}{2 \hbar} \int_{-\pi}^{\pi} \dfrac{\mathrm{d}k_x}{2\pi} \sum_{n \ne m} \dfrac{F(\varepsilon_n) - F(\varepsilon_m)}{\varepsilon_n - \varepsilon_m} \mathrm{Re} \big\{\langle \bm r|n\rangle \langle m|\bm r\rangle (v_{k_x}  (\hat y-y_{\bm r}))_{nm}\big\}\nonumber  \\
    &+ \dfrac{e}{2 \hbar}\int_{-\pi}^{\pi} \dfrac{\mathrm{d}k_x}{2\pi} \sum_{n \ne m} ( f(\varepsilon_n)+  \dfrac{ F(\varepsilon_n) - F(\varepsilon_m)}{\varepsilon_{n}-\varepsilon_{m}}) \mathrm{Re}\big\{|\langle \bm r|n\rangle |^2y_{nm}v_{mn,k_x}\big\}\nonumber \\
    &- \dfrac{e}{2 \hbar}\int_{-\pi}^{\pi} \dfrac{\mathrm{d}k_x}{2\pi} \sum_{n \ne m} ( f(\varepsilon_n) +  \dfrac{ F(\varepsilon_n) - F(\varepsilon_m)}{\varepsilon_{n}-\varepsilon_{m}}) \mathrm{Re}\big\{\langle \bm r|m\rangle \langle n|\bm r\rangle y_{mn}v_{nn,k_x}\big\} \\
    &-\dfrac{e}{2 \hbar}\int_{-\pi}^{\pi} \dfrac{\mathrm{d}k_x}{2\pi} \sum_{n \ne m \ne l}\dfrac{(\varepsilon_{m} - \varepsilon_{l}) F(\varepsilon_n) + (\varepsilon_{l} - \varepsilon_{n})F(\varepsilon_m) + (\varepsilon_{n} - \varepsilon_{m}) F(\varepsilon_l) }{(\varepsilon_{n} - \varepsilon_{l})(\varepsilon_{m} - \varepsilon_{l})} \mathrm{Re}\big\{\langle \bm r|n\rangle \langle l|\bm r\rangle y_{nm}v_{ml,k_x}\big\}  \nonumber
\end{align} We know that $v_{k_x}$ and $\hat y$ are hermitian matrices, so that we have $y_{nm}^* = y_{mn}$ and similarly for $v_{k_x}$. Using the relation $\mathrm{Re} (z) =\mathrm{Re} (z^*)  $ for any complex number $z$, we have
\begin{align}
    &M^z_{\text{orb}}(\bm r) = \dfrac{e}{2 \hbar} \int_{-\pi}^{\pi} \dfrac{\mathrm{d}k_x}{2\pi} \sum_n  f(\varepsilon_n) \mathrm{Re} \big\{ |\langle \bm r|n\rangle|^2 [(v_{k_x}(\hat y-y_{\bm r}))_{nn} + \sum_{m\ne n} y_{nm}v_{mn,k_x}]  - \sum_{m\ne n} \langle \bm r|n\rangle \langle m|\bm r\rangle v_{nn,k_x}y_{nm} \big\} \nonumber\\
    &-\dfrac{e}{2 \hbar} \int_{-\pi}^{\pi} \dfrac{\mathrm{d}k_x}{2\pi} \sum_{n \ne m} \dfrac{F(\varepsilon_n)}{\varepsilon_n - \varepsilon_m} \mathrm{Re} \big\{\langle \bm r|n\rangle \langle m|\bm r\rangle (v_{k_x}(\hat y-y_{\bm r}) + (\hat y-y_{\bm r})v_{k_x})_{nm}\big\}  \\
    & + \dfrac{e}{2 \hbar}\int_{-\pi}^{\pi} \dfrac{\mathrm{d}k_x}{2\pi} \sum_{n \ne m}  \dfrac{ F(\varepsilon_n)}{\varepsilon_{n}-\varepsilon_{m}} \mathrm{Re}\big\{|\langle \bm r|n\rangle |^2y_{nm}v_{mn,k_x} + |\langle \bm r|m\rangle |^2y_{mn}v_{nm,k_x} - \langle \bm r|n\rangle \langle m|\bm r\rangle y_{nm} (v_{nn,k_x} + v_{mm,k_x})\big\} \nonumber \\
    &+\dfrac{e}{2 \hbar}\int_{-\pi}^{\pi} \dfrac{\mathrm{d}k_x}{2\pi} \sum_{n \ne m \ne l}\dfrac{ F(\varepsilon_n)  }{\varepsilon_{n} - \varepsilon_{m}} \mathrm{Re}\big\{\langle \bm r|l\rangle \langle m|\bm r\rangle y_{ln}v_{nm,k_x}+ \langle \bm r|l\rangle \langle n|\bm r\rangle y_{lm}v_{mn,k_x} - \langle \bm r|n\rangle \langle m|\bm r\rangle (y_{nl}v_{lm,k_x}+v_{nl,k_x}y_{lm})  \big\}  \nonumber
\end{align}  In the first and last lines, we have explicitly \begin{align}
    \sum_{m\ne n} y_{nm}v_{mn,k_x} &= (\hat yv_{k_x})_{nn} - y_{nn}v_{nn,k_x}\\
    \sum_{l \ne n,m} y_{nl}v_{lm,k_x}&= (\hat yv_{k_x})_{nm} - y_{nn}v_{nm,k_x} -y_{nm}v_{mm,k_x}  \\
    \sum_{l \ne n,m} v_{nl,k_x}y_{lm} &=  (v_{k_x}\hat y)_{nm} - v_{nn,k_x}y_{nm} - v_{nm,k_x}y_{mm} \\
    \sum_{l \ne n,m} \langle \bm r|l\rangle y_{ln} &= \langle \bm r|n\rangle  y_{\bm r} - \langle \bm r|n\rangle y_{nn}- \langle \bm r|m\rangle y_{mn} \\
    \sum_{l \ne n,m} \langle \bm r|l\rangle y_{lm} &=   \langle \bm r|m\rangle  y_{\bm r} - \langle \bm r|n\rangle y_{nm} - \langle \bm r|m\rangle y_{mm}
\end{align} We can  simplify the first line of this expression by noticing that \begin{align}
    - \dfrac{1}{2}\sum_{m\ne n} \mathrm{Re} \big\{\langle \bm r|n\rangle \langle m|\bm r\rangle &v_{nn,k_x}y_{nm} \big\} = - \dfrac{1}{2}\sum_{m\ne n} \mathrm{Re} \big\{\langle \bm r|n\rangle \langle n|y|m\rangle \langle m|\bm r\rangle \partial_{k_x}\varepsilon_n\big\}\\
    &= - \dfrac{1}{2}\mathrm{Re} \big\{\langle \bm r|n\rangle \langle n|y (1 - |n\rangle \langle n|) |\bm r\rangle \partial_{k_x}\varepsilon_n\big\}\\
    &= - \dfrac{1}{2}\mathrm{Re} \big\{ |\langle \bm r|n\rangle|^2 y_{\bm r} \partial_{k_x}\varepsilon_n\big\}+ \dfrac{1}{2}\mathrm{Re} \big\{ |\langle \bm r|n\rangle|^2 y_{nn} \partial_{k_x}\varepsilon_n\big\}\\
\end{align} which gives us 
\begin{align}
    \hspace*{-.5cm} m_n(\bm r) &= \dfrac{1}{2}  |\langle \bm r|n\rangle|^2  \mathrm{Re} \big\{  (v_{k_x}\hat y + \hat yv_{k_x})_{nn} -2 \partial_{k_x}\varepsilon_n y_{\bm r}\big\} 
\end{align}
The local Berry curvature given by \begin{align}
    \Omega_n(\bm r) =  - \mathrm{Re}\sum_{m \ne n} \dfrac{\langle \bm r|n\rangle \langle m|\bm r\rangle}{\varepsilon_n - \varepsilon_m} ((v_{k_x}\hat y + \hat yv_{k_x})_{nm} - 2v_{nm,k_x}y_{\bm r} )  
\end{align}
We deduce that the local orbital magnetization in the ribbon geometry can be written as \begin{align}
    \Aboxed{M^z_{\text{orb}}(\bm r) =  \dfrac{e}{\hbar} \int_{-\pi}^{\pi} \dfrac{\mathrm{d}k_x}{2\pi} \sum_n \left[  f(\varepsilon_n) m_n(\bm r) + F(\varepsilon_n) \Omega_n(\bm r) \right]}
    \label{eq:Morbribbon}
\end{align}
with \begin{align}
    m_n(\bm r) &=  |\langle \bm r|n\rangle|^2  \mathrm{Re}   (\hat v_{x}(\hat y-y_{\bm r}))_{nn}\\
    \Omega_n(\bm r) &=  - \mathrm{Re}\sum_{m \ne n} \dfrac{\langle \bm r|n\rangle \langle m|\bm r\rangle}{\varepsilon_n - \varepsilon_m} (\hat v_{x}(\hat y-y_{\bm r})+ (\hat y-y_{\bm r})\hat v_{x})_{nm}  
\end{align} where we recall that the velocity operator is given by $\partial_{k_x} \mathcal{H} = v_{k_x}$. Correspondingly, the total orbital magnetization in the ribbon geometry is obtained as  \begin{align}
    M^z_{\text{orb}} =  \dfrac{e}{L\hbar} \int_{-\pi}^{\pi} \dfrac{\mathrm{d}k_x}{2\pi} \sum_n \left[  f(\varepsilon_n) m_n + F(\varepsilon_n) \Omega_n \right]
\end{align} where \begin{align}
    m_n &= \mathrm{Re}   (\hat v_{x}(\hat y-y_{nn}))_{nn}\\
    \Omega_n &=  2\mathrm{Re}\sum_{m \ne n}  \dfrac{ v_{nm,k_x}y_{mn}}{\varepsilon_n - \varepsilon_m}  
\end{align} are respectively the orbital momentum  and the Berry curvature of state $|n,k_x\rangle$.

\bibliographystyle{unsrtnat}
\bibliography{BiblioPaper}
\addcontentsline{toc}{section}{References}